\documentclass[aps,pre,twocolumn,showpacs,superscriptaddress,amsmath,amssymb]{revtex4-1}

\usepackage[english]{babel} 
\usepackage{hyperref}
\usepackage{graphicx}
\usepackage{amsmath}
\usepackage{amssymb}
\usepackage{breakurl}
\newcommand{\rem}[1]{}

\begin{document}

\title{Sliding states of a soft-colloid cluster crystal:
  Cluster versus single-particle hopping}

\author{Mirko Rossini}
\affiliation{Dipartimento di Fisica, Universit\`a degli Studi di Milano,
  via Celoria 16, 20133 Milano, Italy}

\author{Lorenzo Consonni}
\affiliation{Dipartimento di Fisica, Universit\`a degli Studi di Milano,
  via Celoria 16, 20133 Milano, Italy}

\author{Andrea Stenco}
\affiliation{Dipartimento di Fisica, Universit\`a degli Studi di Milano,
  via Celoria 16, 20133 Milano, Italy}

\author{Luciano Reatto}
\affiliation{Dipartimento di Fisica, Universit\`a degli Studi di Milano,
  via Celoria 16, 20133 Milano, Italy}

\author{Nicola Manini}
\email{nicola.manini@fisica.unimi.it}
\affiliation{Dipartimento di Fisica, Universit\`a degli Studi di Milano,
  via Celoria 16, 20133 Milano, Italy}

\begin{abstract}
We study a two-dimensional model for interacting colloidal particles
which displays spontaneous clustering.
Within this model we investigate the competition between the pinning to a
periodic corrugation potential, and a sideways constant pulling force which
would promote a sliding state.
For a few sample particle densities and amplitudes of the periodic
corrugation potential we investigate the depinning from the statically
pinned to the dynamically sliding regime.
This sliding state exhibits the competition between a dynamics where entire
clusters are pulled from a minimum to the next and a dynamics where single
colloids or smaller groups leave a cluster and move across the corrugation
energy barrier to join the next cluster downstream in the force direction.
Both kinds of sliding states can occur either coherently across the entire
sample, or asynchronously: the two regimes result in different average
mobilities.
Finite temperature tends to destroy separate sliding regimes, generating a
smoother dependence of the mobility on the driving force.
\end{abstract}

\maketitle

\section{Introduction}

The standard atomistic approach to tribology, i.e. the study on friction
dissipation and wear, usually focuses on indestructible ``crushproof''
particles, typically atoms or molecules \cite{Muser03, Muser07,
  VanossiRMP13, Vigentini14, Manini15, Manini16, Manini17, Apostoli17,
  Panizon18}.
Recent works \cite{Bohlein12, Hasnain13, Hasnain14, Brazda17, Brazda18}
have brought colloids into the realm of tribology, by letting repulsive
hard-core particles interact with a periodic "corrugation potential",
generated by means of optical forces, and driven by viscous drag.
The conditions of that experiment match the single indestructible particle
paradigm.
Colloids and, more generally, soft matter systems would however allow one
to investigate situations where complex objects, carrying an internal
structure, under the stress produced by external interactions and driving,
can alter their internal structure both in shape and even in the number of
the component sub-particles.
Thus one can expect the presence of a varied set of regimes of response of
the system to the external perturbation.
In the present work, we address precisely this last situation by means of
molecular-dynamics (MD) simulations of a system forming micro phases.
Micro phase formation can take place under a number of conditions on the
interparticle interactions and on the state variables in three dimensions
\cite{Sears99, Sciortino04, Mladek06, Archer07, Archer08}
as well as in two
\cite{Seul95, Imperio04, Imperio06, Diaz-Mendez15}.
The microphases can be disordered
\cite{Sears99, Archer07, Imperio04, Imperio06}
as well as ordered in a crystalline-like state or other patterned
structures like stripes and lamellae
\cite{Leibler80, Fredrickson89, Imperio04, Imperio06, Mladek06, Diaz-Mendez15}
or even as an ordered bicontinous state
\cite{Pini15}
in which both components of a two-component system span the space.
Examples of micro-phase forming systems are block-polymers
\cite{Leibler80, Fredrickson89}
or hard colloids with competing interactions, e.g. short range
attraction-long range repulsion like for some colloid-polymer mixtures
\cite{Seul95, Sears99, Archer07, Imperio04, Sciortino04, Imperio06}.
%
%
Certain soft-matter systems like star polymers or dendrimers in a good
solvent can interpenetrate each other to a large extent so that the
effective interaction between the centers of mass of two such entities has
a soft-core character.
Also such systems can form microphases in which the colloids spontaneously
aggregate into clusters, and the clusters are organized in a crystalline
state \cite{LoVerso06, Mladek08}.
Rheological properties of cluster fluid phase \cite{Imperio08} and of
cluster crystal phase \cite{Nikoubashman11} have been investigated but no
study has yet addressed the tribology of such systems.

In the present work, we simulate a system of interacting colloidal
particles, whose mutual interaction potential generates the {\em
  spontaneous} formation of clusters.
In particular, we concentrate on a 2D geometry with repulsive
interactions, like in many experiments carried out with colloids
\cite{Mangold03, Mikhael08, Pertsinidis08, Mikhael10, Bohlein12,
  Bohlein12PRL}, but the concepts introduced here may be relevant for
vortexes in superconductors \cite{Meng17} too.
%
%
The novelty compared to these experiments is the adoption of an
interparticle interaction which supports the spontaneous formation of
cluster phases.
Cluster dynamics has been investigated in the past \cite{Reichhardt12,
  Reichhardt09b, Reichhardt09c}: in these works clustering was forced by an
external potential, while in the present work clustering is the intrinsic
result of the particle-particle interaction, and is retained even when any
external potentials are turned off.

Section~\ref{model:sec} introduces and motivates the model.
As a first step, in Sec.~\ref{gs:sec} we characterize the $T=0$ equilibrium
states of the model free from external interactions.
We study the ground-state energy of several regular arrangements of
clusters formed by $n$ particles (with $n$ from 1 to 6, that we will refer
to as $n$-clusters) as a function of the mean colloids density, obtaining a
$T=0$ phase diagram.
Then Sect.~\ref{driven:sec} reports the investigation of the dynamics of
the resulting arrays of clusters interacting with a lattice-matched
external periodic potential and an homogeneous pulling force.
Specifically, we report the mobility curves for these cluster systems, as a
function of the intensity of the pulling force, for a few values of the
amplitude of the corrugation potential.
We characterize in detail the different sliding regimes, with entire
clusters advancing, or only parts of them.
The main outcome of this research is that the clusters internal dynamics is
indeed affecting significantly the depinning force and the overall
mobility.

\section{The Model}\label{model:sec}

\subsection{The Interparticle Interaction}\label{hamilt:sec}

\begin{figure}
\centerline{
\includegraphics[width=0.45 \textwidth,clip=]{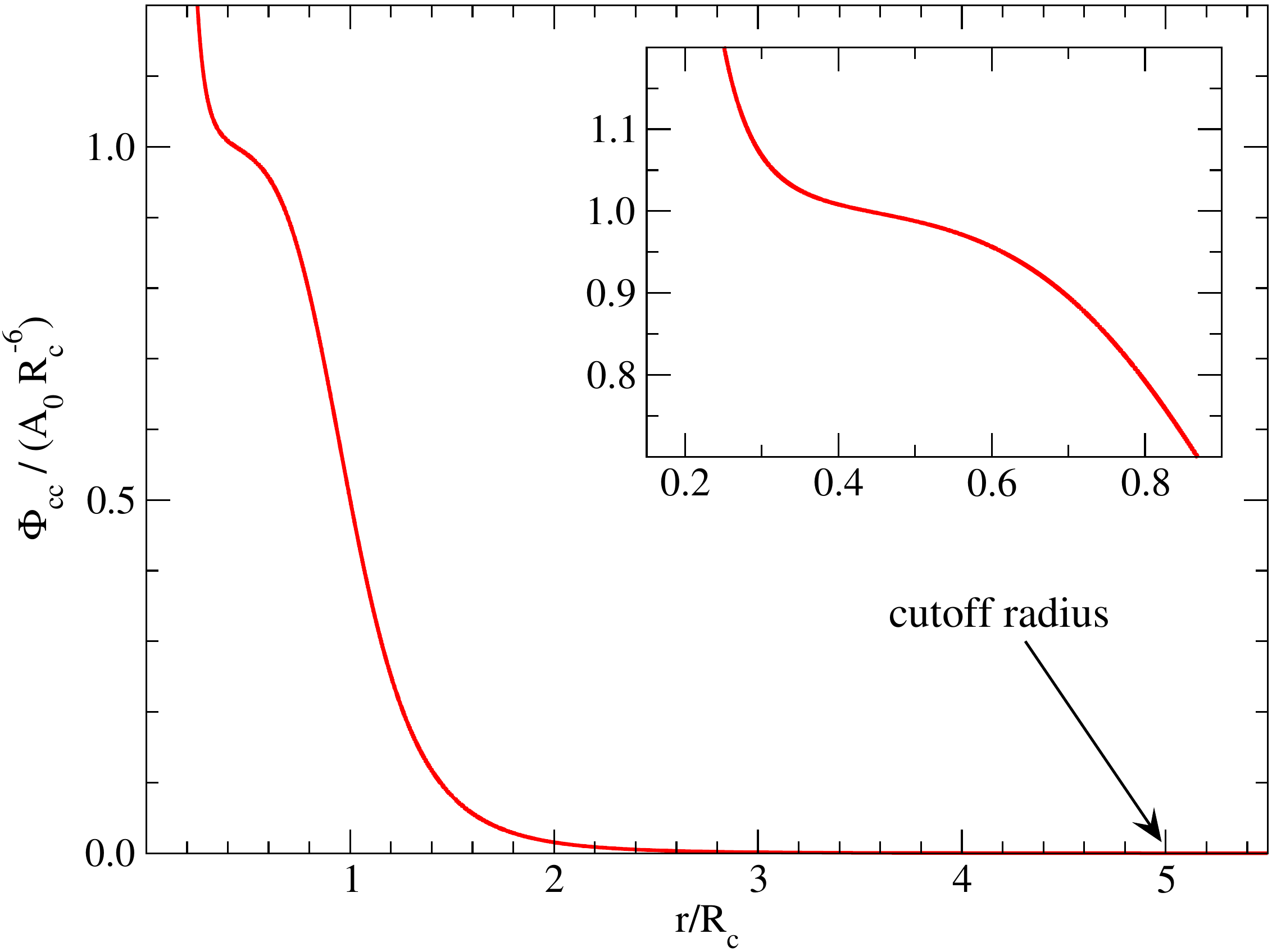}
}
\caption{\label{inter_pot:fig}
  The adopted colloid-colloid potential $\phi_{\rm cc}(r)$,
  Eq.~\eqref{Inter_particle}, consisting on a soft-core function plus a
  weak short-range hard-core term that we introduced to prevent the
  collapse of clusters at $T=0$.
  Inset: detail of the short-distance plateau.
}
\end{figure}

Spontaneous cluster formation can be the result of competitive interactions
\cite{Seul95, Sears99, Archer07, Imperio04, Sciortino04, Imperio06}, but
clusters can also form due to soft potentials
\cite{Diaz-Mendez15,LoVerso06, Mladek08}.
Here we adopt an interaction of the latter type.
We consider the following form for the pairwise interaction potential
energy:
\begin{equation}\label{Inter_particle}
  \phi_{\rm cc}(r) = \frac{A_0}{r^6 + R_c^6} + \frac{B_0}{r^6}
\,.
\end{equation}
This interaction, depicted in Fig.~\ref{inter_pot:fig}, consists of two
terms: a soft-core repulsive interaction, and a weak hard-core repulsion.
The $B_0=0$ (purely soft-core) model has been fully characterized in two
dimensions in Ref.~\cite{Diaz-Mendez15}: its phase diagram consists in a
high-temperature low-density regular fluid phase, and a low-temperature
high-density cluster phase.
The reason for the cluster phase is that when the inter-particle distance
decreases under $r\lesssim R_c$, the potential energy $\phi_{\rm cc}(r)$
flattens out to a nearly constant energy, with the result of producing a
quite small repulsive force.
As a result, in one dimension a particle in between two other particles
kept at fixed positions, when the distance between these fixed particles is
not too large, may be better off energetically by coming closer to either
of them, rather than remaining at the middle point.
%
This suggests the mechanism whereby, in any dimensionality and at large
enough density, a spontaneous symmetry breaking leading to clustering can
be energetically favorable against a uniform fluid phase.
A detailed mathematical criterion has been developed for the formation of
cluster phases in the case of soft-core potentials in terms of their
Fourier transform becoming negative in a range of finite wave vector
\cite{Likos01}.
At high $T$, entropic effects and fluctuations tend to favor the uniform
fluid phase against cluster formation.
In the opposite $T\to 0$ extreme, for $B_0=0$ no mechanism keeps the
colloids apart, and so the clustering tendency would reach the extreme
limit of collapsed point-like clusters with all colloids of a cluster
sitting exactly at the same position.
To prevent this singular behavior we add the hard-core $B_0$ term of
Eq.~\eqref{Inter_particle} for whose coefficient we adopt a relatively
small value $B_0 = 5\times10^{-5}\,A_0$ \cite{Consonni16} in order to
perturb the phase diagram as little as possible, at least in the
not-too-high-density region that we are interested in.

\begin{table}
  \begin{center}
    \begin {tabular}{l|l}
      \hline
      \hline
      Physical quantity & System units \\
      \hline
      Length & $R_c$ \\
      Energy & $E_0=A_0/R_c^6$ \\
      Mass & $m$ \\
      \hline
      Number density & $R_c^{-2}$ \\
      Force & $F_0=A_0/R_c^7$ \\
      Time & $t_0 = R_c^4\sqrt{m/A_0}$ \\
      Velocity & $v_0=R_c^{-3}\sqrt{A_0/m}$ \\
      Viscous damping & $\eta_0 = R_c^{-4}\sqrt{A_0m}$ \\
      Mobility & $\mu_0 = R_c^4/\sqrt{A_0m}$ \\
      \hline
      \hline
    \end{tabular}
  \end{center}
  \caption{\label{units:tab}
    The system of units adopted in the simulations.
    Every physical quantity is expressed in terms of these natural model units.
  }
\end{table}

We take the repulsive-potential characteristic distance $R_c$, and
characteristic energy $A_0/R_c^6$, and the mass $m$ of the colloidal
particles as fundamental units.
In the following we express all physical quantities in terms of suitable
model units, as listed in Table~\ref{units:tab}.

\subsection{The Equations of Motion}\label{eq:sec}

In simulations, we let the entire system evolve following the standard
Langevin dynamics in two dimensions \cite{Gardiner}, under the effect of an
external periodic corrugation potential $U_{\rm ext}(\mathbf r)$ and a
constant driving force $F$ applied to each particle.
The equation of motion for the $j^{\rm th}$ particle is:
\begin{eqnarray} \label{langevin:eq}
  m\ddot{\mathbf r}_j &=&
  F \, \hat{\mathbf x} - \eta \dot{\mathbf r}_j +
  \\\nonumber
 && - {\mathbf \nabla}_{{\mathbf r}_j} \left[ \sum\limits_{k \ne j}^N
    \phi_{\rm cc}(|{\mathbf r}_k-{\mathbf r}_j|) + U_{\rm ext}(\mathbf
    r_j) \right] + {\mathbf \xi}_j \,.
\end{eqnarray}
Here $\eta$ is the coefficient of viscous friction associated to the
fluid where the colloids move; we generate an overdamped dynamics by
adopting a large $\eta = 28\, \eta_0$.
%
%
The ${\mathbf\xi}_j$ terms are Gaussian-distributed random forces with
amplitude $\sigma = \sqrt{2m\eta k_{\rm B}T/\Delta t}$, where $\Delta t$
is the simulation time step: they simulate the collisions of the
molecules in the fluid with the colloids, thus generating the
appropriate Brownian dynamics.
Together, the viscous and the random-force terms in Eq.~\eqref{langevin:eq}
represent a standard Langevin thermostat \cite{Gardiner}.
In simulations, the inter-particle potential $\phi_{\rm cc}$ is cutoffed
smoothly (vanishing potential and potential derivative) at a distance
$R_{\rm cutoff} = 5\,R_c$ \cite{Consonni16}.

\begin{figure}
\centerline{
\includegraphics[width=0.45 \textwidth,clip=]{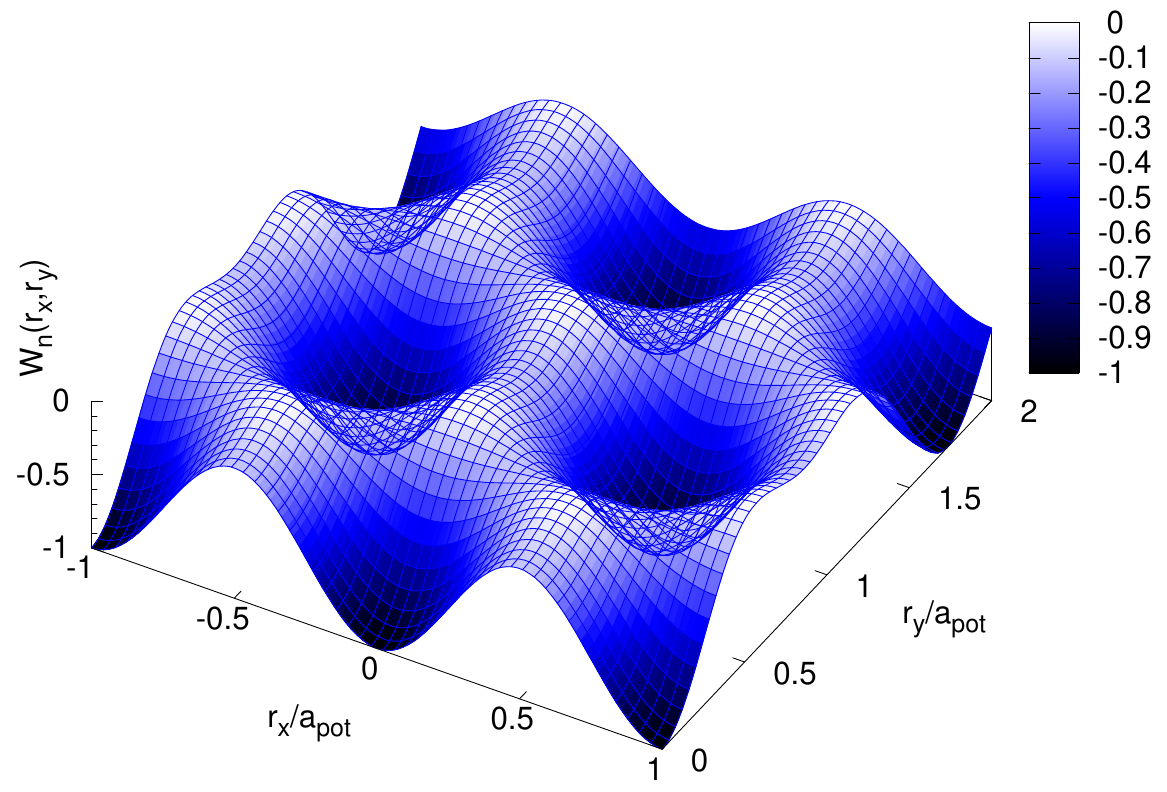}
}
\caption{\label{ext_potential:fig}
  The profile $W(r_x,r_y)$ of the corrugation potential which tends to
  localize the colloids.
  It is assumed to take a periodic hexagonal-lattice form, with spacing
  $a_{\rm pot}$, see Eq.~\eqref{Wfunc:eq}.
}
\end{figure}

The periodic external potential $U_{\rm ext}(\mathbf r)$ simulates the
effects of friction against a crystalline surface.
In the laboratory, this interaction has been realized by means of a
modulated light field constructed by laser interference, so that its
spacing and intensity that can be tuned with considerable freedom
\cite{Bohlein12}.
In our simulations, we assume a corrugation potential of the form:
\begin{equation}
 U_{\rm  ext}({\mathbf r}) = V_0\,W({\mathbf r})
 \,.
\end{equation}
Here $V_0$ is the amplitude of the spatial modulation and $W(\vec{r})$ is a
2D periodic function with hexagonal symmetry, period $a_{\rm pot}$, and
unit amplitude.
Its explicit form is:
\begin{eqnarray}\nonumber
  W({\mathbf r}) &=&
  -\frac{2}{9} \left[ \frac{3}{2}
    + 2\cos\left(\frac{2\pi r_x}{a_{\rm pot}}\right)
    \cos\left(\frac{2\pi  r_y}{\sqrt{3}a_{\rm pot}}\right) \right.
    \\ \label{Wfunc:eq} & &
    \left.
    + \cos\left(\frac{4\pi r_y}{\sqrt{3}a_{\rm pot}}\right)
    \right]
  .
\end{eqnarray}
This function is illustrated in Fig.~\ref{ext_potential:fig}.

Following the $r_y = 0$ path in the $x-$direction a particle goes
through alternating minima and saddle points of $W({\mathbf r})$.
This is the energetically less costly trajectory that a colloid can
follow when pushed by an $x$-directed force.
Along this path, the external potential is a simple sinusoidal
oscillation
\begin{equation}
  U_{\rm ext}(r_x,0) =
  -V_0\,\left[
    \frac{5}{9} + \frac{4}{9}\cos\left(\frac{2\pi r_x}{a_{\rm pot}}\right)
    \right]
  \,.
\end{equation}
The resulting energy barrier separating the saddle points from the
minima is therefore $\frac{8}{9} \,V_0$.
In this situation the static friction of an isolated particle at $T =
0$, i.e.\ the minimum force needed to push that particle across the
potential corrugation when inertial effects are negligible, equals the
largest value of $\frac{\partial}{\partial r_x}U_{\rm ext}(r_x, 0)$,
namely:
\begin{equation}\label{F1s}
  F_{1s} = \frac{8\pi V_0}{9a_{\rm pot}}
\,.
\end{equation}
In the simulations and in the relative graphs we express all applied
forces by comparison to the static friction $F_{1s}$ of the single
colloid, expressed in Eq.~\eqref{F1s}.
For the lattice spacing we take $a_{\rm pot} = 1.5\, R_c$, and select
values of the particles density that favor arrays of clusters whose
spacing matches precisely this separation, as discussed in
Sect.~\ref{cell:sec}.

\subsection{The Cell and the Initial Condition}\label{cell:sec}

We simulate the 2D model in a parallelogram-shaped supercell with
periodic boundary conditions (PBC), thus fixing the average density.
To respect the hexagonal symmetry that the arrays repulsive colloids (or
clusters thereof) tend to adopt, we use a supercell generated by two
primitive vectors of equal length $L$ and forming an angle of $60^\circ$
with each other.
For $V_0=0$ we are free to adopt an arbitrary cell size: in the static
simulations of Sect.~\ref{gs:sec} we will vary $L$ freely in order to
evaluate the energetics as a function of the colloid density, and therefore
determine the $T=0$ phase diagram of the free model.

For $V_0\neq 0$, this supercell needs to match the symmetry of the
corrugation potential: we must stick to $L$ given by an integer multiple of
$a_{\rm pot}$.
We settle for a compromise size $L=12\,a_{\rm pot}$, small enough to
guarantee simulations involving a not-too-large number of particles (and
therefore manageable simulation times), but large enough for a fair
averaging over fluctuations and allowing for independent and possibly
asynchronous movements of individual clusters.
Accordingly, the cell contains $144$ corrugation minima in an area
$A=\sqrt{3} L^2/2$.

As discussed below, we find that a cluster-cluster spacing $1.5 R_c$ as
optimal for favoring a well-defined $n$-cluster phase.
In the present work we focus on a lattice-matched condition between the
cluster-cluster spacing and the spacing $a_{\rm pot}$ between adjacent
minima of the corrugation.
Therefore in all dynamical simulations of Sect.~\ref{driven:sec} we
settle for fixing $a_{\rm pot}=1.5\, R_c$, and therefore a supercell
side $L=18\,R_c$, thus generating a supercell area $A\simeq 280.6\,
R_c^2$.
In this supercell, the potential has $144$ wells, each initially hosting
a single particle or a $n = 2$ to $6$-particles cluster.
We simulate a total $N=144\, n$ particles, which thus sample the
following discrete values for the average density:
\begin{equation} \label{rhoval:eq}
  \rho = \frac NA =
  n\,\frac 2{\sqrt{3} \, a_{\rm pot}^2} =
  n\, \frac 8 { 9 \sqrt{3} \, R_c^2}
\,.
\end{equation}

For the initial condition, we start off with distances $\delta = 0.32\,
R_c$ between the colloids within each cluster suggested by the inflection
of the inter-particle potential, Fig.~\ref{inter_pot:fig}.
This initial configuration relaxes very quickly to the equilibrium
interparticle distances that depends on the state of the system.
We start with a regular lattice of preformed clusters, because the
alternative possibilities of starting, e.g., with the particles
regularly or randomly distributed across the supercell leads to
irregular patterns involving different cluster sizes, which may describe
appropriately large temperature and/or phase-coexistence conditions, but
surely do not represent the optimal single-type cluster phases at $T=0$.

\section{Ground-state Energetics for the Free Model} \label{gs:sec}

\begin{figure}
\centerline{
\includegraphics[width=0.4 \textwidth,clip=]{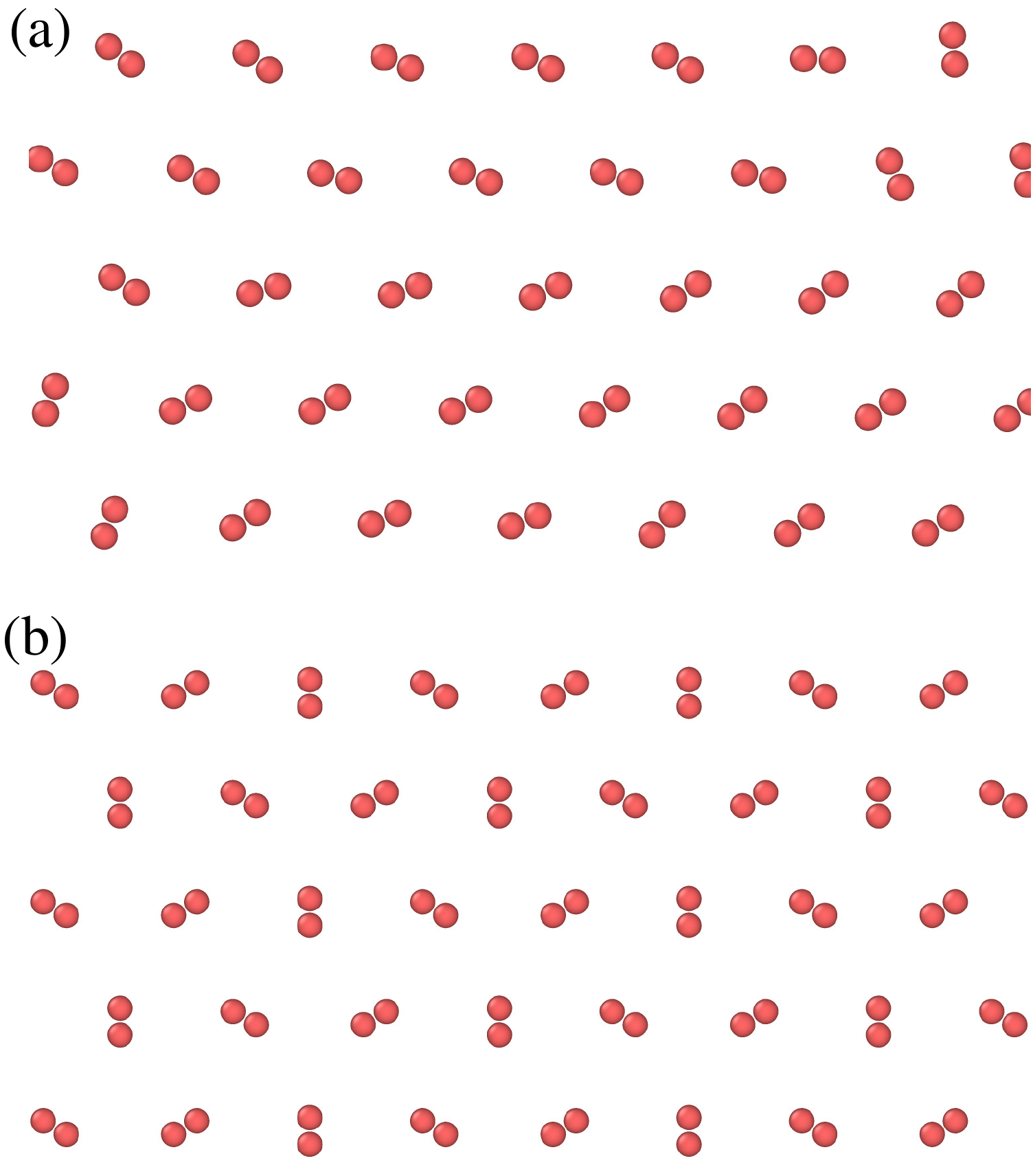}
}
\caption{\label{cluster2_orient_gs:fig}
  Small portions of fully relaxed $T=0$ configurations representing
  examples of alternative local minima for the total repulsive energy of
  the free model ($V_0 = 0$, $F=0$) in a $n=2$ cluster phase at density
  $\rho = 16 /( 9 \sqrt{3} \, R_c^2) \simeq 1.0264\,R_c^{-2}$.
  The repulsive energy per particle of these two configurations is (a)
  $E = 1.108486 \, E_0$; (b) $E =1.108907\, E_0$.
}
\end{figure}

While at finite temperature the number of colloids in each cluster
fluctuates, at $T=0$, for every given density $\rho$, one expects that
an hexagonal lattice of clusters with the same number $n$ of particles
represents a local minimum of the total repulsive energy.
We explore the density range from $\rho= 0.4\,R_{c}^{-2}$ to
$4\,R_{c}^{-2}$.
In the absence of corrugation potential ($V_0=0$) and driving force
($F=0$), for every $n$-cluster and every density (i.e. every lattice
spacing of the cluster crystal), we execute a full relaxation by means
of a damped dynamics, for at least a total time $2000\, t_0$, but
extending the relaxation for longer time until the residual kinetic
energy is safely below $10^{-8}\, E_0$.
This method leads us very close to the precise ground state.
Depending on $n$ and $\rho$, the system may remain trapped in a local
minimum characterized by some residual energy associated to the
cluster's orientational order, see Fig.~\ref{cluster2_orient_gs:fig}.
We verified that these orientational-order energies are quite small, of
the order of $10^{-4} E_0$ per particle.
For this reason and for the irrelevance (discussed below) to the driven
dynamics, we do not investigate the details of the orientational order
in the ground state of the clustered phases.

\begin{figure}
\centerline{
\includegraphics[width=0.45 \textwidth,clip=]{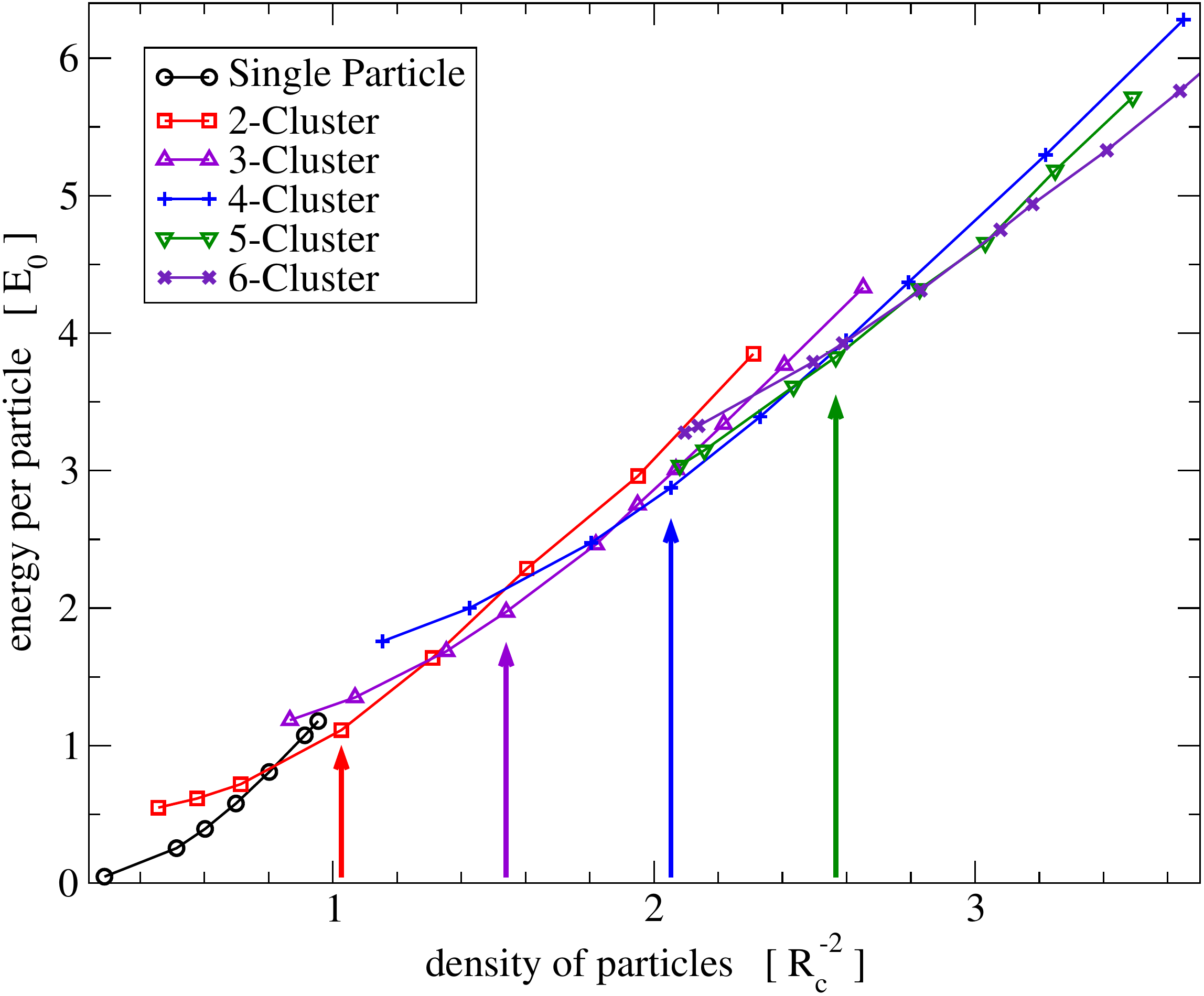}
}
\caption{\label{ground_state:fig}
  The static ($T=0$, $F=0$) free ($V_0=0$) model total colloid-colloid
  repulsive energy per particle, as a function of the number density
  $\rho$ of colloids.
  Individual curves compare different $n$-cluster configurations,
  $n=1,\dots 6$.
  The orientational order is irrelevant here, as orientational-energy
  differences are well within the size of the reported points, see
  Fig.~\ref{cluster2_orient_gs:fig}.
  Arrows identify the densities adopted for the driven dynamics studied in
  Sect.~\ref{driven:sec}.
}
\end{figure}

By running several energy minimizations starting from different cluster
phases, we can follow the evolution of the ground-state energy as a
function of the density $\rho$.
For each configuration we evaluate the total colloid-colloid repulsive
energy, divide it by the number of colloids in the supercell, and obtain
the energy per particle.
Figure~\ref{ground_state:fig} displays the curves of the total energy
per particle for each $n=1$ to $6$, as a function of $\rho$.
We see that at each given density, several cluster configurations
represent competitive local energy minima.
However, at each density one of them is usually lowest in energy.
Crossings between successive curves identify the coexistence density
values where the $n$-cluster and the $(n+1)$-cluster have the same
stability, and can therefore coexist.
By carrying out a finer numerical analysis and a Maxwell construction
one could also identify density ranges characterized by the phase
coexistence of homogeneous domains consisting of clusters of different
$n$.
We need not go into this detail: based on the obtained rough phase diagram,
we identify the densities corresponding to stable $n$-cluster states,
characterized by a cluster-cluster spacing $1.5 R_c$, and we adopt them for
the driven simulations of Sect.~\ref{driven:sec}.

\section{Driven Dynamics} \label{driven:sec}

\begin{figure}
  \centerline{
    \includegraphics[width=0.45 \textwidth,clip=]{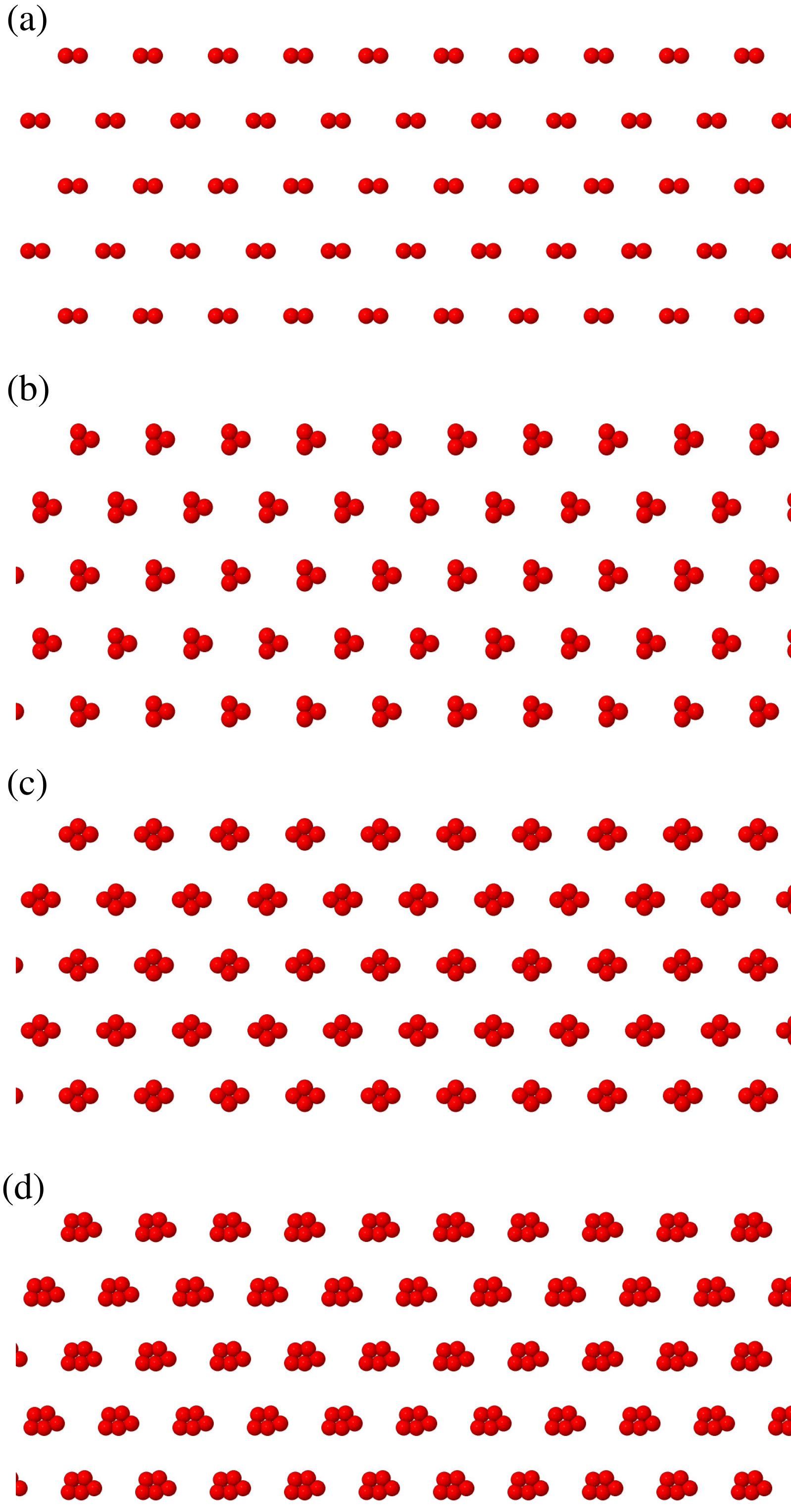}
  }
\caption{\label{cluster_pulled:fig}
   Small portions of the $T=0$ configurations in a $V_0 = 1.5\,E_0$
   corrugation, under the action of a $F=0.6\, F_{1s} < F_{\rm dep}$
   pulling force, i.e.\ these are static configurations prior to
   depinning, for:
  (a): $n=2$; (b): $n=3$; (c): $n=4$; (d): $n=5$ clusters.
}
\end{figure}

Starting from the equilibrated state obtained as discussed in
Sect.~\ref{gs:sec}, we investigate the dynamics under the competing
effects of the external periodic corrugation potential, which tends to
immobilize the clusters at its minima, and the lateral driving force $F$
which, if there was no corrugation, would tend to establish a sliding
state at speed $v=F/\eta$.
As expected of a commensurate, and even lattice-matched configuration, a
static pinning threshold is always present.
When $F$ is slowly raised from $0$ up to this pinning threshold, after an
initial transient allowing for the cluster rearrangement, the steady
sliding speed remains null.
However, nontrivial rearrangements of the clusters are observed well in
advance of depinning:
the driving force clears all orientational orderings characteristic of
the ground states.
The clusters are forced to specific orientations, which are displayed in
Fig.~\ref{cluster_pulled:fig}.
These ferro-orientational \cite{Ozaki99,Potocnik12} arrangements are made
energetically advantageous by the anisotropic energy landscape present at
the regions in between the minimum and the saddle point of the corrugation
potential of Fig.~\ref{ext_potential:fig} where $F$ pushes the clusters.
Precisely this force-induced ferro-orientational cluster arrangement
observed for all $n$ prior to depinning \cite{Reichhardt09a}
makes the orientational details of the free-model ground state
(e.g.\ Fig.~\ref{cluster2_orient_gs:fig}) irrelevant for the dynamics we
address in the present work.

\begin{figure}
\centerline{
\includegraphics[width=0.45 \textwidth,clip=]{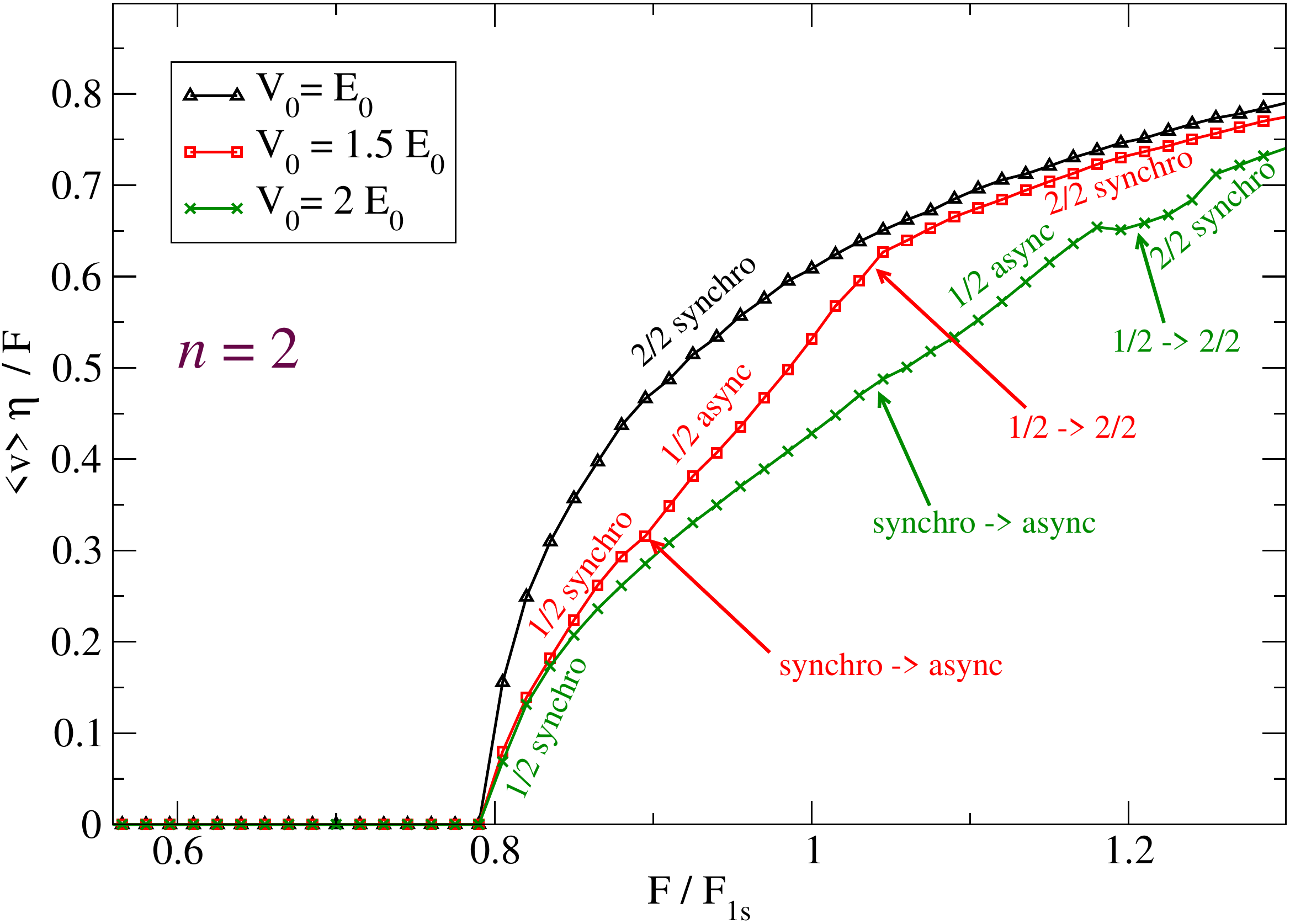}
}
\caption{\label{mob_cl2:fig}
  The $T=0$ mobility of the $n=2$-cluster lattice as a function of the
  driving force $F$ expressed as a fraction of the depinning force
  $F_{1s} = 8\pi V_0/(9a_{\rm pot})$ for an isolated particle, for
  corrugation amplitudes $V_0/E_0 = 1$, $1.5$, and $2$.
  The numerical error affecting the mobility due to averaging over
  noninteger numbers of periods is less than 0.3\% of the free mobility
  $\eta^{-1}$.
  The labels along the curve identify different sliding regimes, with
  groups of $m/n$ particles leaving one well of $U_{\rm ext}(\mathbf r)$,
  and moving to the next well at the right.
  Arrows identify transitions from a sliding regime to the one occurring at
  larger force.
  {\it Synchro}/{\it async} regimes refer to whether all well-to-well
  particle jumps occur simultaneously across the cell or they do not.
  In this specific $n=2$ case, even without specification, all motions
  are ``1D'', meaning that colloids advance along horizontal lines
  through the minima of $W(\mathbf r)$.
}
\end{figure}

\begin{figure}
\centerline{
\includegraphics[width=0.45 \textwidth,clip=]{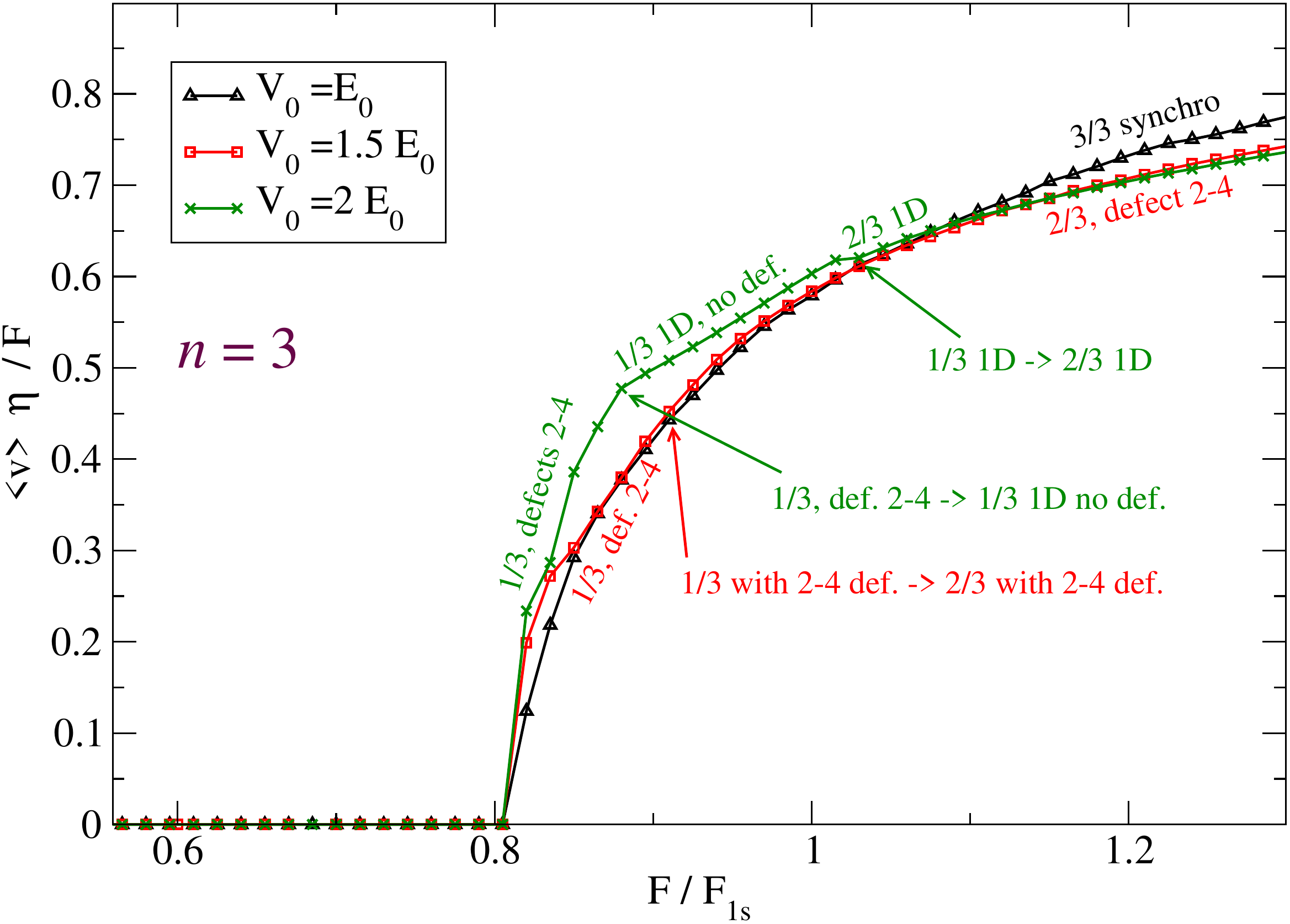}
}
\caption{\label{mob_cl3:fig}
  Same as Fig.~\ref{mob_cl2:fig} but for the $n=3$ cluster lattice.
}
\end{figure}

\begin{figure}
\centerline{
\includegraphics[width=0.45 \textwidth,clip=]{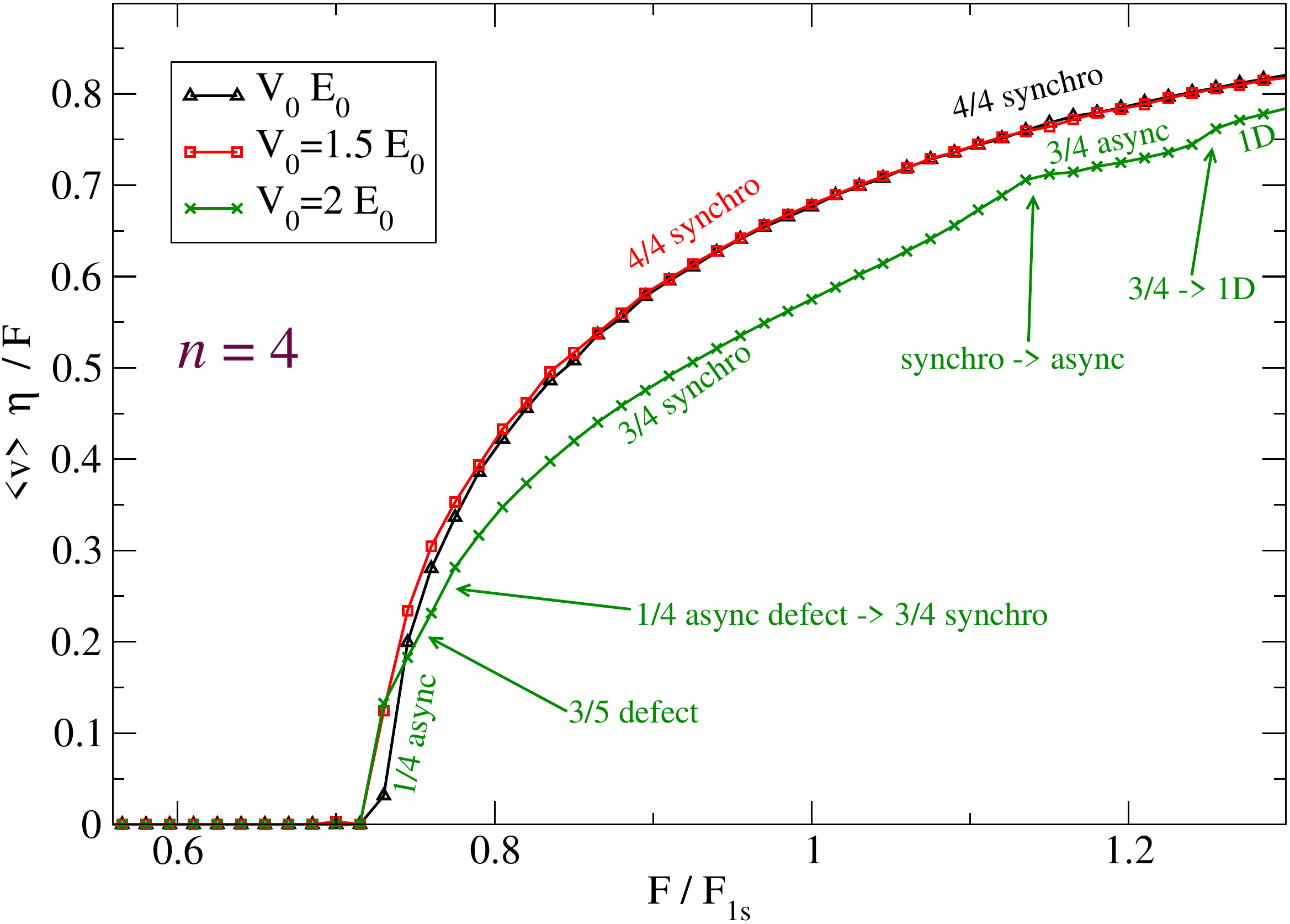}
}
\caption{\label{mob_cl4:fig}
  Same as Fig.~\ref{mob_cl2:fig} but for the $n=4$ cluster lattice.
}
\end{figure}

\begin{figure}
  \centerline{
    \includegraphics[width=0.45 \textwidth,clip=]{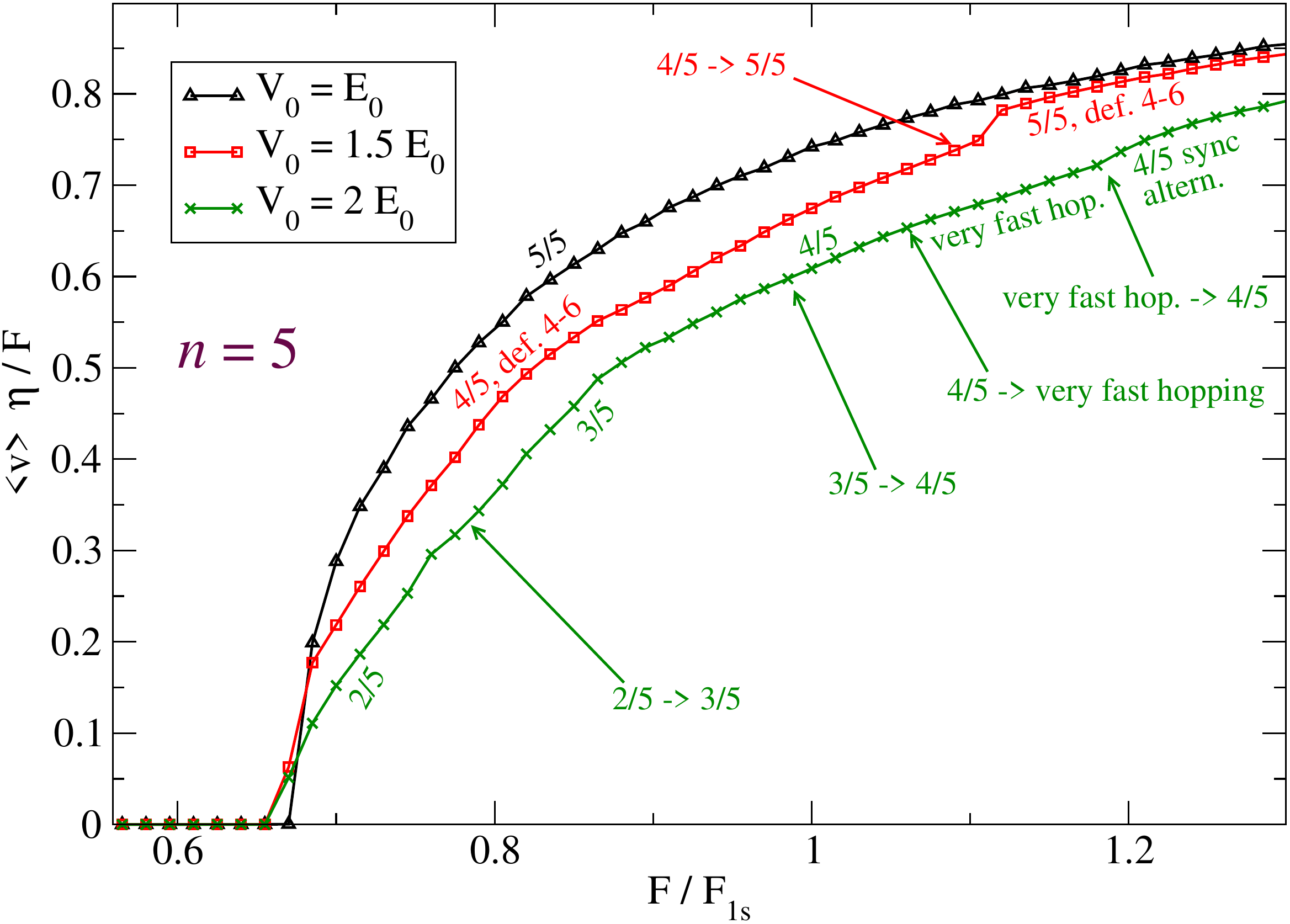}
  }
\caption{\label{mob_cl5:fig}
  Same as Fig.~\ref{mob_cl2:fig} but for the $n=5$ cluster lattice.
  All observed dynamical modes are synchronized here.
  See text for a description of the {\it very fast hop} and {\it $4/5$
    sync altern} regimes.
}
\end{figure}

The most straightforward dynamical indicator for a driven model of this
kind is the average mobility, i.e. the ratio of the average
$\mathbf{\hat x}$-directed velocity to the driving force, which we
report for different numbers $n$ of colloids per cluster.
If the corrugation potential was removed ($V_0=0$), then for any nonzero
force, the mobility would equal the free mobility $\eta^{-1}$, which
equals $0.0357\, \eta_0$ with the present choice of the damping
parameter.
The addition of the corrugation leads to smaller values of mobility.

To evaluate the mobility, for each $2\leq n\leq 5$ we execute three
sequences of simulations, with the following amplitudes of the periodic
external potential: $V_{0} = E_0$, $1.5\,E_0$ and $2\,E_0$.
For each sequence, we increase the applied lateral force in a range
proportional to $V_0$, namely $0.55\,F_{1s} \leq  F \leq 1.3\, F_{1s}$ in
small steps $\Delta F=0.015\,F_{1s}$.
At each step, the average velocity is evaluated by averaging over a total
time $1400\, t_0$ after an initial transient time $600\, t_0$.
Figures~\ref{mob_cl2:fig}--\ref{mob_cl5:fig} report the resulting
mobilities of the cluster lattices for the investigated numbers $n$ of
colloids per cluster as a function of the applied driving force $F$.
In the initial small-$F$ simulations the driving force is insufficient to
extract the colloid clusters from the potential wells of $U_{\rm ext}$:
thus a null mobility is obtained.
Beyond a certain static friction threshold $F_{\rm dep}$, the colloids
unpin and start to advance, resulting in a finite mobility, which grows
progressively approaching the free-particle value $\eta^{-1}$.
The small ``adiabatic''increase in $F$ allows us to evaluate $F_{\rm
  dep}$ with a relatively small uncertainty, and to identify the
features of the particle motion as mobility increases after depinning.

\begin{figure}
\centerline{
\includegraphics[width=0.45 \textwidth,clip=]{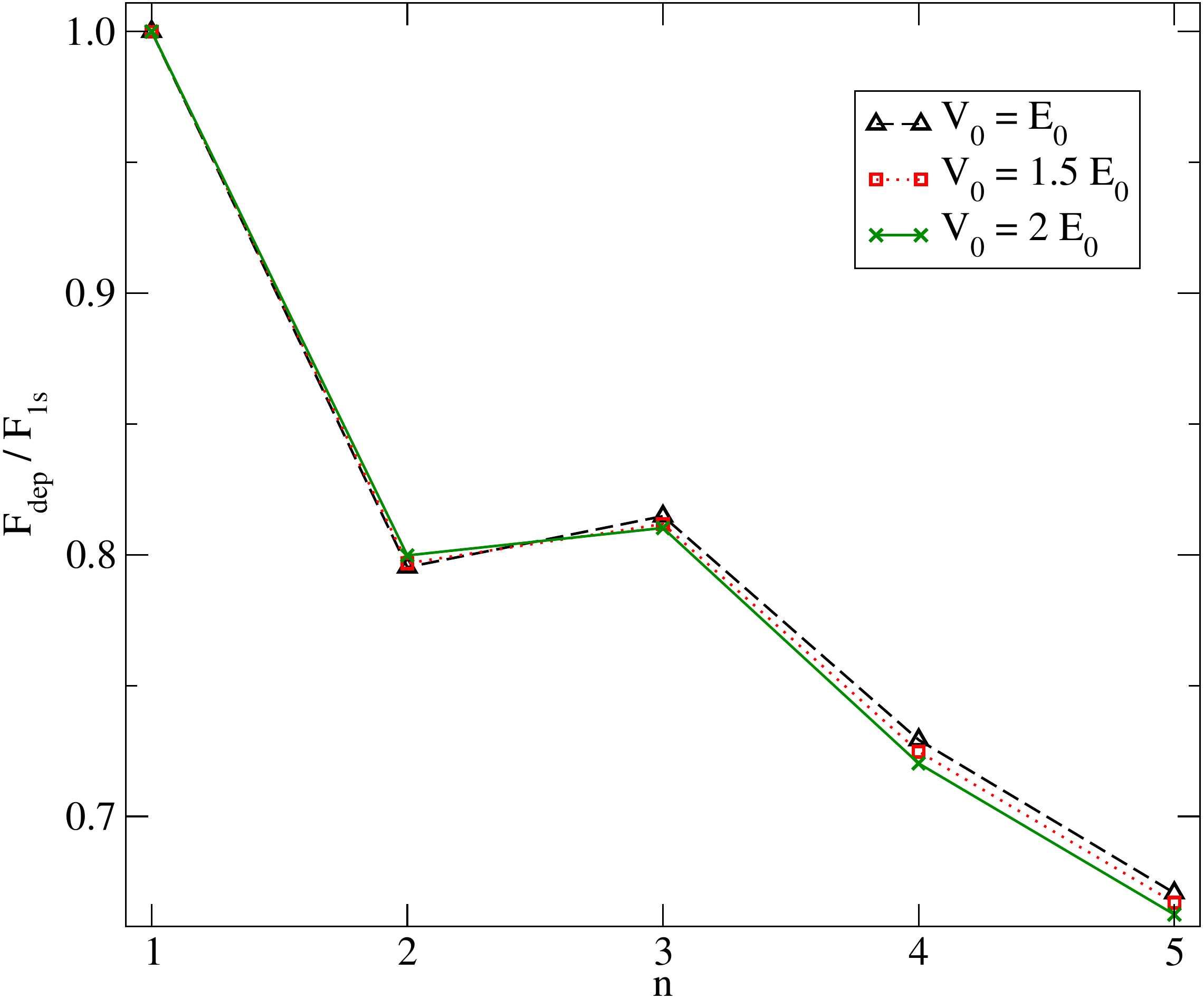}
}
\caption{\label{dep_force:fig}
  The static friction, or depinning threshold, $F_{\rm dep}$, expressed as
  a fraction of the single-colloid depinning force $F_{1s}$ (also relevant
  for the depinning of the $n=1$ no-cluster phase), for $V_0/E_0 = 1$,
  $1.5$, and $2$, as a function of the number $n$ of colloids per cluster.
}
\end{figure}

Consider the depinning force $F_{\rm dep}$.
Although its values can be read directly from
Figs.~\ref{mob_cl2:fig}--\ref{mob_cl5:fig}, for ease of comparison
Fig.~\ref{dep_force:fig} reports them as a function of $n$.
A first observation is that the clustering of the colloids causes a strong
decrease of $F_{\rm dep}$ compared to the value $F_{\rm dep}=F_{1s}$ of a
single colloid, by about 20\% for $n=2$ and $3$ up to 33\% for $n=5$.
The static friction force peaks for $n = 3$.
Above $n = 3$ the depinning force tends to decrease as $n$ increases, as
expected due to the growing size of the clusters.

The $n = 2$ clusters mark an exception, as they depin at smaller force
than $n = 3$.
Also, in the comparison of the depinning force for three values of
$V_0/E_0$, we see that depinning for $n = 2$ occurs at a smaller $F_{\rm
  dep}/F_{1s}$ for $V_0 = E_0$ than $V_0 = 2 E_0$, contrary to all other
cluster sizes.
The reason for the special behavior of the $n = 2$ cluster lays in its
elongated conformation under traction, see Fig.~\ref{cluster_pulled:fig}.
As a result, in the pulling direction $\hat {\mathbf x}$ the $n = 2$ cluster
extends over a longer distance $\approx \delta$ than the size $\approx
\delta \sqrt{3}/2$ of $n = 3$.
For this reason, when the cluster center of mass approaches the inflection
point of the corrugation potential (the point beyond which the retaining
force begins to taper off), the rightmost particle has moved forward more
for $n = 2$ than for $n = 3$.

After depinning, the mobility, on top of an overall smooth increase as a
function of $F$, exhibits sharp changes in slope,
Figs.~\ref{mob_cl2:fig} - \ref{mob_cl5:fig}.
These changes mark transitions between different sliding patterns, which
can be examined by monitoring the advancement of the individual colloids.
The following section describes precisely these sliding regimes.

\subsection{Sliding regimes}\label{patterns:sec} 

\begin{figure}
  \centerline{
    \includegraphics[width=0.95 \columnwidth,clip=]{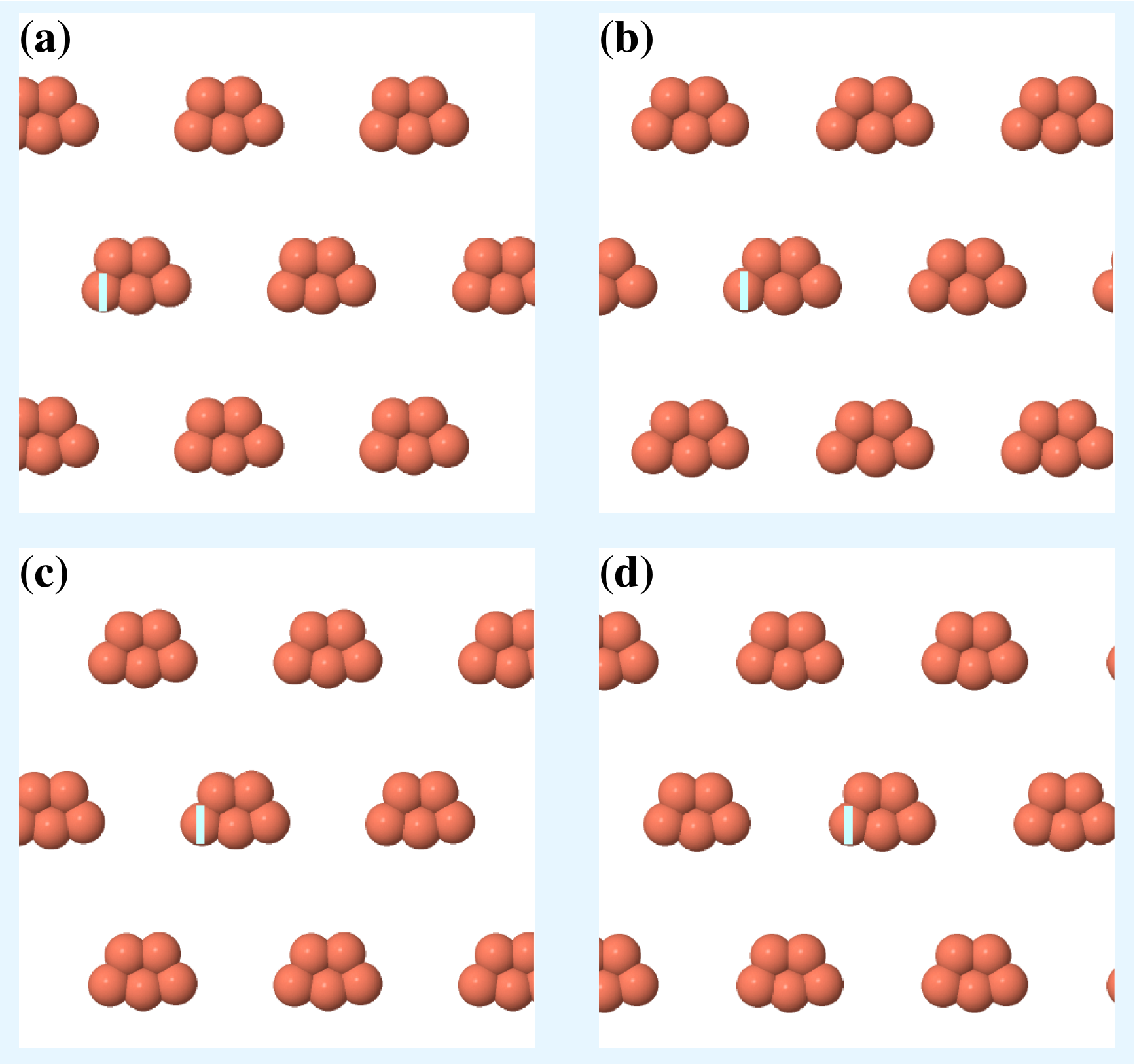}
  }
\caption{\label{5_5sync:fig}
  The whole-cluster ($5/5$) advancement observed in the $n=5$, $V_0 =
  E_0$ simulation driven at $F/F_{1s} = 1$, see Fig.~\ref{mob_cl5:fig}.
  Panels (a) - (d) represent subsequent snapshots, separated by $\Delta
  t = 2\,t_0$.
  A particle is labeled with a vertical bar for ease of identification
  in the successive snapshots.
}
\end{figure}

Under driving, the internal structure of the clusters makes them advance
either as a single object, or less trivially with structural
decompositions and recompositions.
Depending on the corrugation amplitude $V_0$ and on the number of
colloid per cluster we observe (or not) several of these
decomposition-recomposition phenomena as the lateral force is increased.
For convenience, we indicate the observed dynamical patterns on the
mobility curves of Figs.~\ref{mob_cl2:fig} - \ref{mob_cl5:fig}, where
the switching between patterns can induce visible mobility anomalies.
We use $k/n$ labels indicating dynamical patterns which persist for an
entire force interval.
The $k/n$ labels in the mobility curves indicates that the advancement
from one well to the next occurs with $k$ particles out of a cluster of
$n$ jumping ahead, joining the next cluster, and leaving $n-k$ particles
behind.
Arrows point at transitions from one kind of dynamical behavior to the
next.

\begin{figure} 
  \centerline{
    \includegraphics[width=0.95 \columnwidth,clip=]{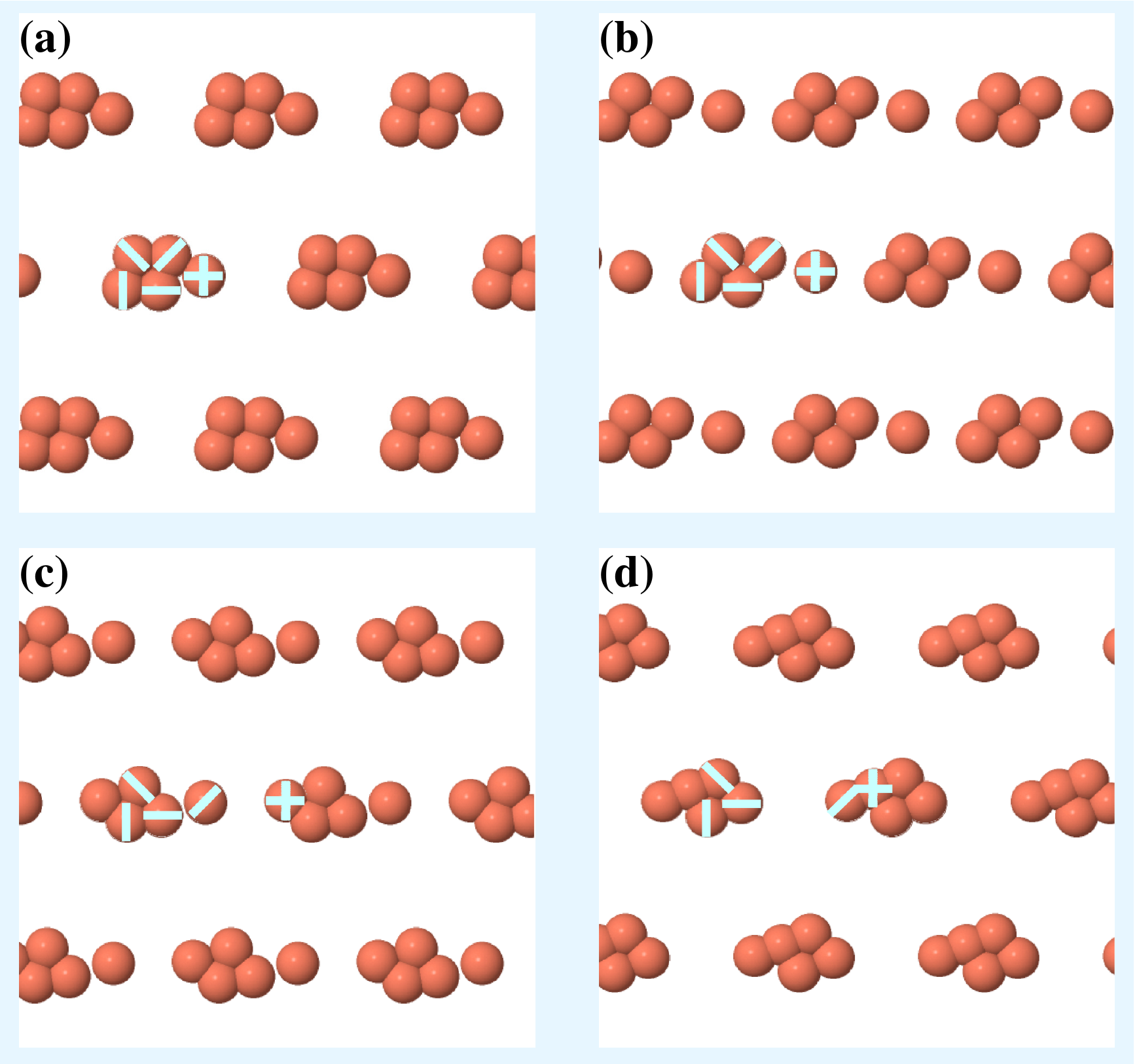}
  }
\caption{\label{2_5sync:fig}
  Successive snapshots illustrating the $2/5$ dynamics observed in the
  $n=5$, $V_0 = 2\, E_0$ simulation driven at $F/F_{1s} = 0.7$, see
  Fig.~\ref{mob_cl5:fig}.
  Five particles are labeled for ease of identification in the
  successive snapshots.
  The time interval $\Delta t = 2\,t_0$, the same as in
  Fig.~\ref{5_5sync:fig}.
}
\end{figure}

\begin{figure}
\centerline{
\includegraphics[width=0.45 \textwidth,clip=]{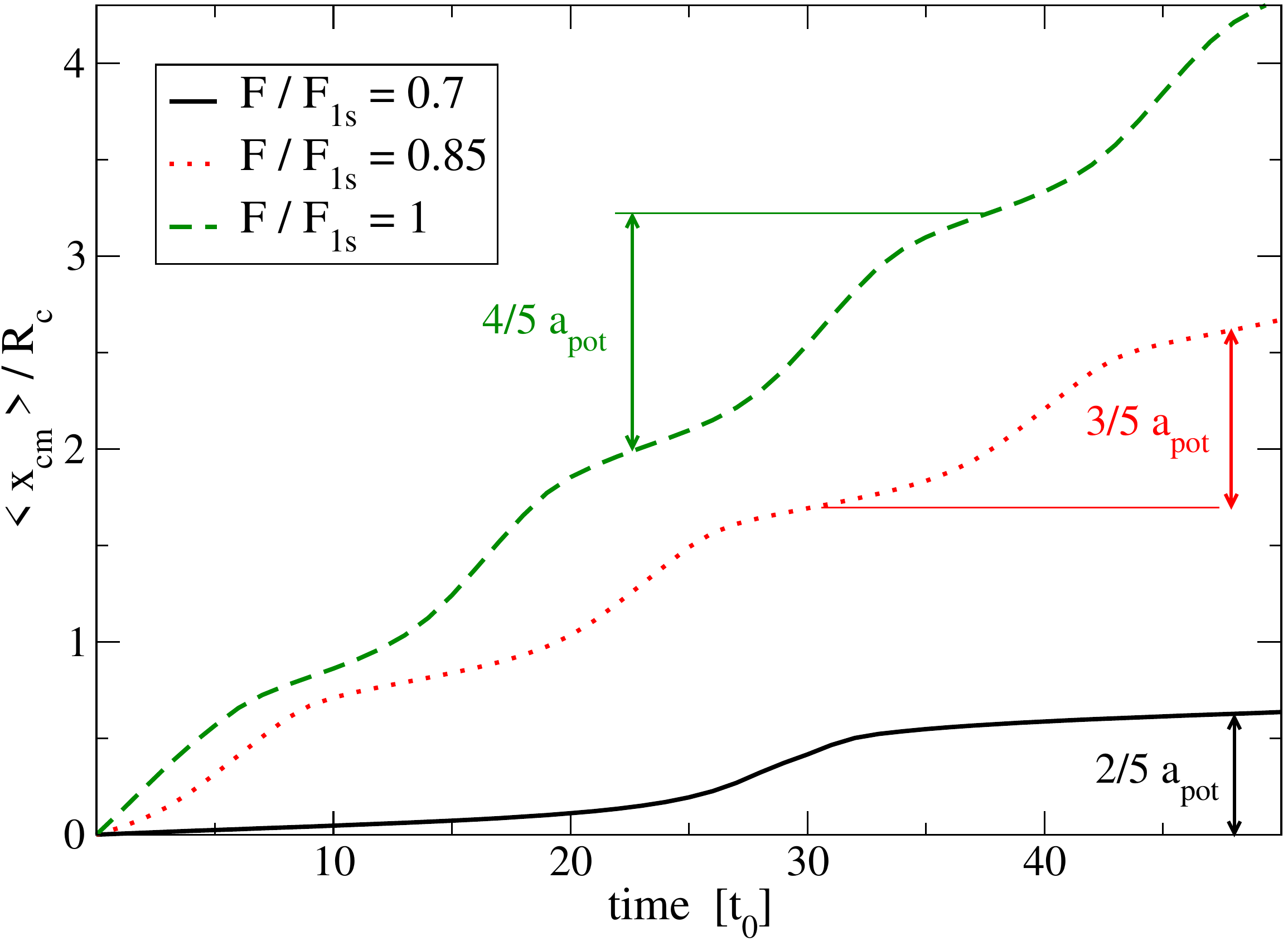}
}
\caption{\label{scalini:fig}
  The time-dependence of the $x_{\rm cm}$ of the system with $n = 5$
  colloids per cluster for $V_0 = 2\, E_0$ and for three different
  forces representative of the fully synchronized $2/5$, $3/5$ and $4/5$
  dynamical modes indicated in Fig.~\ref{mob_cl5:fig}: given $a_{\rm
    pot} = 1.5$, the steps heights are $0.62178 \simeq \frac 25 \,
  a_{\rm pot}$, $0.88585 \simeq \frac 35 \, a_{\rm pot}$ and $1.23788
  \simeq \frac 45 \, a_{\rm pot}$.
}
\end{figure}

As a general rule, a small corrugation amplitude $V_0$ tends to favor the
clusters advancing as a whole, namely in ``$n/n$'' modes.
In contrast, for larger amplitude $V_0$ the corrugation competes
favorably with the energy advantage of the clusters to remain entire,
and as a result for larger $V_0$ we do observe nontrivial
decomposition-recomposition patterns.
These decompositions can occur synchronously across the entire supercell
(and in this case they are labeled {\em synchro}) or asynchronously ({\em
  async} label).
Figures~\ref{5_5sync:fig} and \ref{2_5sync:fig} report successive
snapshots illustrating the two main kinds of dynamical patterns, as
marked on the mobility plots.
Specifically, Fig.~\ref{5_5sync:fig} illustrates the most basic $5/5$
advancement mode: whole cluster advancement from one corrugation well to
the next, as observed for weak corrugation $V_0 = E_0$.
In contrast, Fig.~\ref{2_5sync:fig} illustrates an advancement mode
involving synchronous decompositions and recompositions of the $n=5$
clusters, with $2$ particles abandoning each cluster and jumping to the
next corrugation well while leaving $3$ particles behind at each step.
Examples of similar patterned movements involving different numbers of
particles are reported in the Supplementary Material
\cite{supporting_current}, as digital videos.
Figure~\ref{scalini:fig} illustrates the effect of these patterned
movements on the $x$-coordinate of the center of mass of the colloids.
As the particle hoppings are synchronized across the entire sample, the
center-of-mass position $x_{\rm cm}$ exhibits relatively fast jumps
during the synchronized hoppings, alternating with plateaus during the
clusters rearrangements.
The heights of the observed steps are a fraction of $a_{\rm pot}$, which
matches precisely the specific dynamical ratio, e.g.\ the $2/5$ pattern
(solid line in Fig.~\ref{scalini:fig}) exhibits $\frac 25\, a_{\rm
  pot}$-high steps.

When the driving force reaches relatively large values $F\gtrsim F_{\rm
  1s}$, its action on the system can force the collective dynamics into
an effectively one-dimensional (1D) dynamics, with all colloids
organized in rows parallel to the direction of the driving force
($\hat{x}$).
We noted this mode on the mobility plots, using the appropriate {\it 1D}
label, except for Fig.~\ref{mob_cl2:fig}, because for $n=2$ the motion
is quasi-1D for any $F$.
Similar quasi-1D dynamics was observed also in the different model of
Ref.~\cite{Reichhardt12, Reichhardt09a, Reichhardt09c}.
In contrast for $n = 5$, this quasi-1D advancement does not develop
completely, because of a significant transverse ($\hat{\mathbf
  y}$-directed) displacement persisting at all times due to the
hard-core repulsion not leaving enough room for 5 colloids in a row of
length $a_{\rm pot}$.
For $n = 5$, the large-$F$ dynamics can consist of a rapid and
uninterrupted transfer of particles from one cluster to the next, as
identified by the label {\it very fast hop} in Fig.~\ref{mob_cl5:fig}.
Alternatively we have observed a {\it 4/5 sync altern} motion, with groups
of 4 colloids jumping across clusters arranged in an alternating pattern (3
above + 2 below) -- (2 above + 3 below) along the rows, as illustrated in
video {\tt cl5\_4.44V2.avi} in the Supplementary Material
\cite{supporting_current}.

The synchronization of the well-to-well jumps can be promoted or
disfavored by weak cluster-cluster interactions.
As a consequence, we occasionally observe a de-synchronization of the
well-to-well hoppings for different clusters.
In the Supplementary Material \cite{supporting_current} we report examples
of asynchronous dynamics, where the advancements at different rows occur at
subsequent times.
Contrasted to the synchronous advancement steps of Fig.~\ref{scalini:fig},
in asynchronous dynamics as a consequence of averaging no (or very small)
steps are observed in the overall center-of-mass coordinate.
The asynchronous jumps can occur in a regular sequence at different rows or
with apparently chaotic hoppings as for example in the video
{\tt cl4\_2.71V2.avi} provided as Supplementary Material
\cite{supporting_current}.
Note that any regularity in asynchronous motions (and possibly the
overall synchronization itself when present) is favored by the specific
supercell periodic boundary conditions adopted in simulation: as such,
the synchronized or regular patterns may be considered artifacts of the
simulated model.

This zero-temperature dynamics, often dominated by unique regular or
chaotic attractors \cite{Alligood00,Ott02} does not represent a
realistic condition that one could put in relation to actual
experiments.
Indeed, in the more realistic finite-temperature simulations of
Sect.~\ref{finiteT:sec} below, irregular asynchronous depinning occurs
in most cases, due to the Brownian forces randomly anticipating or
retarding the advancement of individual particles or clusters.

By cycling the driving force $F$ up and down, we investigated whether these
dynamical systems retain some kind of dynamics memory, i.e. the sliding
state is affected not only by the competition of the applied confinement
potential and driving, but also by the initial sliding configuration.
Indeed in these $T=0$ simulation we do find some small difference of the
boundaries between specific dynamical modes between the decreasing-$F$ path
and the increasing-$F$ path.
To rule out non-physical underdamping effects, we verified that such
hysteretic effects persist even under a doubled damping rate $\eta = 56
\,\eta_0$, indicating that a competition among multiple dynamical
attractors is indeed present in this model.

In the present study the adopted $\hat {\mathbf x}$-directed orientation
of the driving force tends to drive the colloids straight across the
saddle points of the corrugation potential.
By exploring other more general pushing directions, depending on the
angle formed by the force with the $\hat{\mathbf x}$-direction, the
colloids would be driven toward the corrugation maxima too.
This kind of investigation has been carried out in the past for
different models or experimental setups, typically finding locking to
energetically and/or dynamically favorable advancement directions
\cite{Reichhardt04, Reichhardt11, Bohlein12PRL}.
We expect similar directional-locking phenomena in the present model.
However, possible novel regimes realized by the current cluster model
may include concurrent multiple decomposition paths for the clusters,
with several particles leaving simultaneously a potential well to reach
different neighboring ones.
We leave the investigation of these regimes to future research.

\subsection{Thermal Effects}\label{finiteT:sec}

\begin{figure}
\centerline{
\includegraphics[width=0.36 \textwidth,clip=]{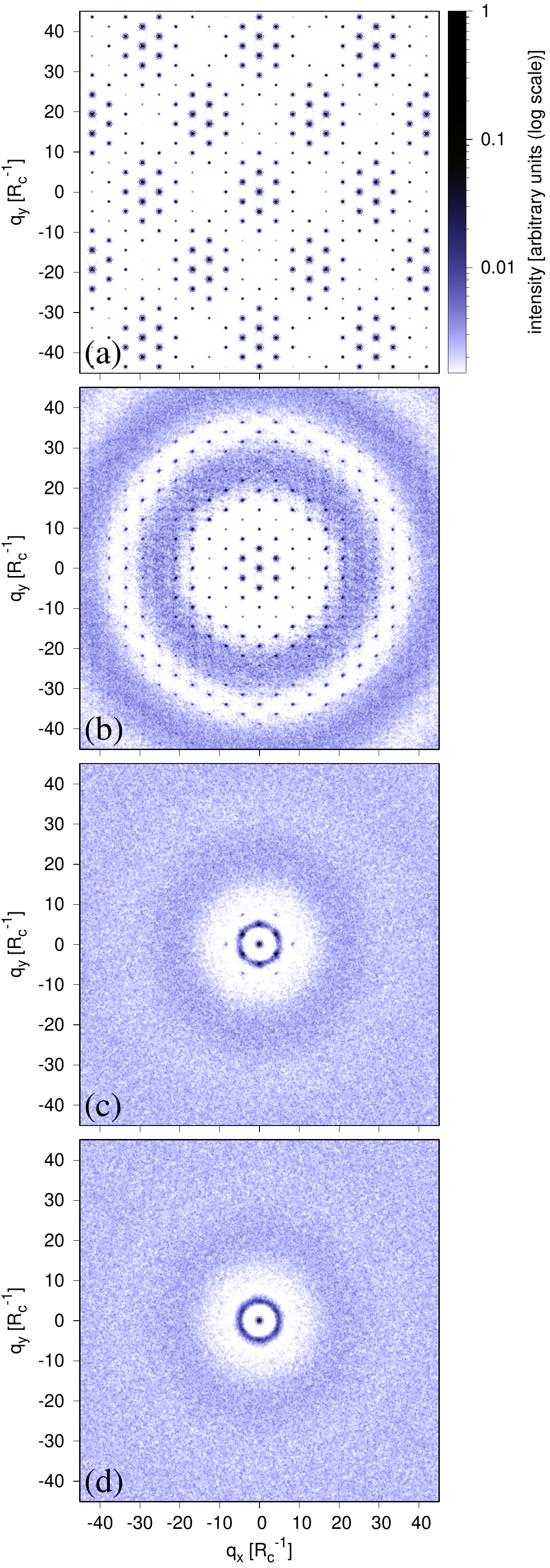}
}
\caption{\label{structure:fig}
  The scattering intensity $S({\mathbf q})$ obtained assuming that each
  colloid acts as structureless point-like scatterer.
%
  The simulations refer to $n=4$, at four temperatures expressed in units
  of energy (i.e.\ assuming $k_{\rm B}=1$): (a) $T=0$ -- the ordered state
  in the static condition depicted in Fig.~\ref{cluster_pulled:fig}c, (b)
  $T=0.01\,E_0$ -- rotational melting of the cluster lattice, (c)
  $T=0.15\,E_0$ -- crystal lattice phase just below structural melting, and
  (d) $T=0.17\,E_0$ -- a liquid phase just above melting.
  The $T>0$ simulations of panels (b-d) are carried out for the free model
  ($V_0 = 0$, $F=0$).
  $S({\mathbf q})$ is averaged over 30 snapshots widely spaced in time
  along an equilibrium simulation.
}
\end{figure}

We simulate finite temperature by restoring the random forces
${\mathbf\xi}_j$ in the equations of motion \eqref{langevin:eq}, and
therefore sampling the canonical ensemble.
The starting point of the $T>0$ simulations is the appropriate $T=0$
static configuration (a perfect cluster lattice) obtained in
Sect.~\ref{gs:sec}.
In the simulated scattering intensity reported in
Fig.~\ref{structure:fig}a, the modulation of the Bragg peaks reflects
the structure factor of the orientationally ordered lozenge-shaped
clusters at $T=0$, see Fig.~\ref{cluster_pulled:fig}c.
In the finite-temperature simulations of the free model ($V_0 = 0$, $F=0$),
it is straightforward to verify that even a low temperature, such as
$T=0.01\, E_0$ induces the melting of the orientational order of the
clusters, as seen in the diffuse rings at integer multiples of $2\pi/\delta
\simeq 20 \, R_c^{-1}$ reported in Fig.~\ref{structure:fig}b.
Further raising temperature leads rapidly to more disordering of the
cluster crystal, and eventually to melting, which occurs between
$T=0.15\, E_0$ and $T=0.17\, E_0$, see Fig.~\ref{structure:fig}c,d.

According to this analysis, we focus the investigation of the driven
model on moderately-low temperature $T\leq 0.05\, E_0$.
In the simulations under driving, as we did in the $T=0$ protocol,
successive runs at larger and larger driving force $F$ are started from the
final configuration of the previous step in $F$, which is increased
``adiabatically'' by $\Delta F = 0.01\,F_{1s}$ at each step.
For each step, the simulation duration is $500\,t_0$: the first 30\% is
dropped to prevent transient effects, and the remainder is used for the
determination of the mobility, providing a fair averaging over thermal
fluctuations.

Strictly speaking, at finite temperature there is no static friction,
because thermal fluctuations would lead to a diffusive displacement of the
layer even for $F=0$, and to a slow systematic drift in the force direction
for $F>0$, if one had the patience to wait long enough for these
thermally-activated rare events to occur.
However, at the relatively small considered temperatures $T\ll V_0$, the
rate of well-to-well thermal-assisted hopping is extremely small, and as
long as the applied force is small $F \ll F_{\rm 1s}$, we observe no such
hoppings at all for the duration of our simulations.
This allows us to define $F_{\rm dep}$ even for $T>0$, as the threshold
force beyond which a significant center-of-mass displacement is observed
in the applied force direction within the simulation duration.
For forces immediately before this threshold, the mobility is not
precisely null, but it is really tiny, with overall advancements of the
entire system by far less than one lattice spacing $a_{\rm pot}$ over
the entire simulation time.
We note also that all particle well-to-well displacements occur within
horizontal rows: although {\it a-priori} possible, we never observe any
particle abandoning their initial $\hat{\mathbf x}$-oriented row and
transferring to a well in the next row.
These cross-row hopping would of course become frequent for larger
$T/V_0$.
We do however observe particle jumps within rows, leading to a limited
but nonzero random concentration of $n-1$ and $n+1$ ``defective''
clusters in the pre-depinning quasi-immobile state.

\begin{figure}
\centerline{
\includegraphics[width=0.45 \textwidth,clip=]{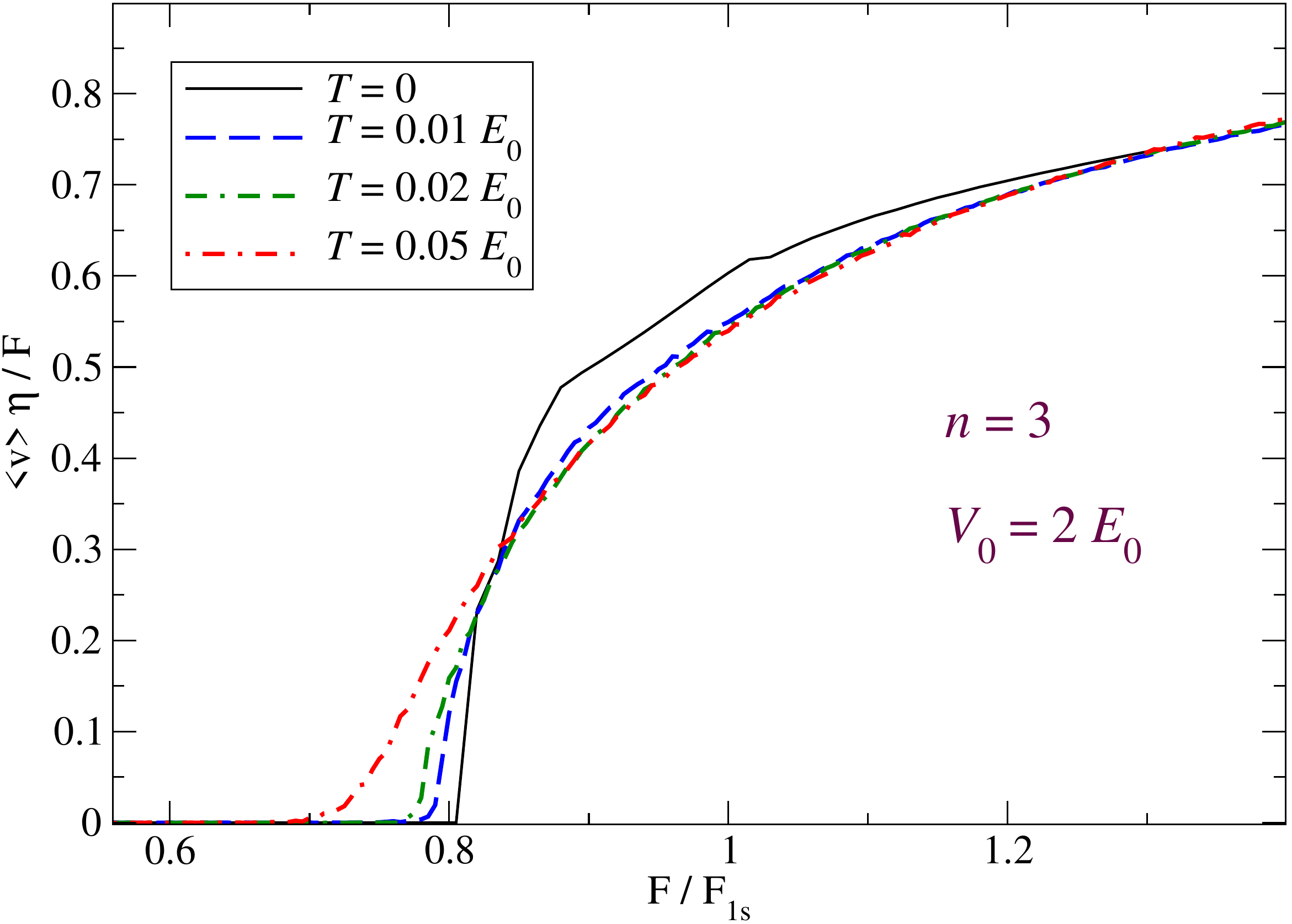}
}
\caption{\label{cl3V2:fig}
  The mobility, relative to the value $\eta^{-1}$ of free colloids, as a
  function of the driving force, relative to the $T=0$ depinning threshold
  $F_{1s}$ of isolated colloids, at three relatively small but finite
  temperatures, expressed in units of energy $E_0$.
  The $T=0$ solid curve is the same as in Fig.~\ref{mob_cl3:fig}.
  Simulations are carried out for $n=3$-particle clusters, for corrugation
  $V_0 = 2 E_0$.
  Statistical fluctuations account for error bars (not drawn) on mobility
  mostly smaller than $5\times 10^{-3} \eta^{-1}$.
}
\end{figure}

The mobility resulting in the finite-temperature simulations is exemplified
in Fig.~\ref{cl3V2:fig}.
This figure exhibits the following characteristic features, common to
analogous simulations that we carried out for clusters of sizes $n=2$ to
$5$, and reported in Appendix:
(i) The depinning threshold $F_{\rm dep}$ decreases for increasing
temperature, because thermal fluctuations help the particles anticipate the
barrier hopping, thus activating sliding even when the driving force is
nominally insufficient to overcome the barrier.
(ii) For $F\gtrsim F_{\rm dep}$, immediately above the threshold, thermal
fluctuations tend to favor the sliding state, as shown by the higher
mobility proving a sort of thermolubric effect \cite{Gnecco00, Sang01,
  Dudko02, Riedo03, Reimann04, Krylov05, Jinesh08, Franchini08, Pierno14}.
However, depending on $n$ and the corrugation amplitude $V_0$, for forces
larger than the $T=0$ depinning threshold $F_{1s}$, increasing temperature
can lead to either a slightly decreasing mobility, as in
Fig.~\ref{cl3V2:fig}, or essentially $T$-independent mobility, quite
similar to the values obtained for $T=0$, see Appendix~\ref{thermal:app}.

The main effect of finite temperature on the cluster dynamics is to
suppress the $T=0$ synchronized advancements, especially those
associated to the fractional-advancement modes.
As a result, at $T>0$ the center-of-mass advancement is generally
smoother, with no, or at least substantially smeared, steps like those
occurring for the $T=0$ fractional sliding regimes reported in
Fig.~\ref{scalini:fig}.
Nonetheless, some amount of approximate synchronization is preserved in the
special case of whole-cluster hopping, e.g.\ at $T>0$ the {\it
  $3/3$~synchro} pattern of Fig.~\ref{mob_cl3:fig} even extends down to
smaller $F$ than for $T=0$.
Relatedly, the sharp transitions between different sliding modes observed
at $T=0$, at finite $T$ are replaced by smooth crossovers.

\section{Discussion and Conclusions}

In the present work we introduce and study a driven model for friction at
the micro scale where the protagonists are spontaneously formed clusters,
rather than simple structureless objects\cite{Thompson90b, Shinjo93,
  Smith96, He99, Muser00, Consoli00, Vanossi00, Muser01, Consoli01,
  Vanossi04a, Guerra10, BraunManini11, ManiniBraun11, vandenEnde12,
  Hasnain14, Varini15, Ma15, Koren16a} or structured but unbreakable
units\cite{molshapeLin, Sivebaek12}, as usually considered in previous
investigations.
We focus on a lattice-matched situation, with the external corrugation
periodicity matching exactly the equilibrium spacing between the clusters.
In the language of the Frenkel-Kontorova model \cite{Floria96, Braun98,
  Braunbook, Popov11, VanossiRMP13}, this is the typical situation which
maximizes the pinning effect of the corrugation, leading to a finite static
friction regardless of how small the corrugation amplitude may be.
We do find indeed a finite static friction in our simulations of the
current model, with clustering affecting the static-friction threshold and
the sliding mechanisms for driving force exceeding this threshold.

For small corrugation amplitude, the clustering tendency is dominating the
dynamics, so that clusters tend to advance as a individual objects, with
the partial exception of single-particle hopping induced by thermal
fluctuations near the depinning threshold.
In contrast, at larger corrugation amplitude, the stability of clusters is
challenged when the driving force moves them in regions where the curvature
of the external corrugation potential is negative: this leads to cluster
decomposition and partial hoppings.
In this regime, the internal structure of the clusters decomposes in
several ways as a function of the driving force, thus producing a quite
rich dynamics with complicate fractional advancement patterns for clusters
of various size.
We investigate the impact of these phenomena on the mobility of the system.
At $T=0$ we observe and characterize well-defined regimes of sliding of
entire clusters or of decomposition-recomposition processes, with
deformation of the clusters.
These processes can take place in a synchronous or in asynchronous way
across the system.
Thermal effects tend to destroy such distinct regimes: we observe a gradual
change of sliding under increasing driving force, already for temperature
as low as $\sim 5$\% of the temperature at which the transition from a
cluster crystal to a uniform fluid occurs.

The phenomenon of cluster decomposition/recomposition under driving is not
expected to be specific of the interparticle interaction adopted here.
We therefore predict that experiments carried out in conditions where
spontaneous clustering occurs should observe this kind of dynamical
behavior.

In the future we plan to extend the investigation to lattice-mismatched
configurations.
In incommensurate conditions, for increasing corrugation amplitude $V_0$
there is room for an Aubry-type transition from a ``superlubric'' dynamics
characterized by "solitonic" sliding modes \cite{Aubry78, Peyrard83,
  Floria96, Braun98, Santoro06, Vanossi07Hyst, Vanossi07PRL, Manini07PRE,
  Cesaratto07, Castelli08Lyon, Castelli09, Vanossi12PNAS, Vanossi13,
  MandelliPRL15, MandelliPRB15, Paronuzzi16, Mandelli17, Bonfanti17}.
A similar condition could be realized by taking for the substrate a
symmetry other than hexagonal, as e.g.\ in Ref.~\cite{Persson93b,
  ManiniBraun11, McDermott13a, McDermott13b}.
In the small-$V_0$ superlubric ground state, a finite fraction of clusters
is supposed to sit near one of the maxima of the incommensurate corrugation
potential, rather than all in a potential well as in the fully commensurate
setup of the present paper.
The novelty for a cluster-supporting condition is that, as $V_0$ is raised,
two competing effects may arise: either an entire cluster near the maximum
slides down toward a nearby potential well, thus leading to a relatively
standard Aubry-type transition \cite{MandelliPRL15,Mandelli17}, or
alternatively it decomposes under the action of divergent corrugation
forces, leading to a structurally inequivalent ground state.
Which of these mechanisms turns in earlier is open for investigation.
These competing possibilities are likely to lead to a rich phase diagram.

\begin{acknowledgments}
  We acknowledge useful discussion with C. Bechinger, R. Guerra, and
  A. Vanossi.  The COST Action MP1303 is also gratefully acknowledged.
\end{acknowledgments}

\appendix

\section{Thermal effects on mobility}
\label{thermal:app}

\begin{figure}
\centerline{
\includegraphics[width=0.7 \columnwidth,clip=]{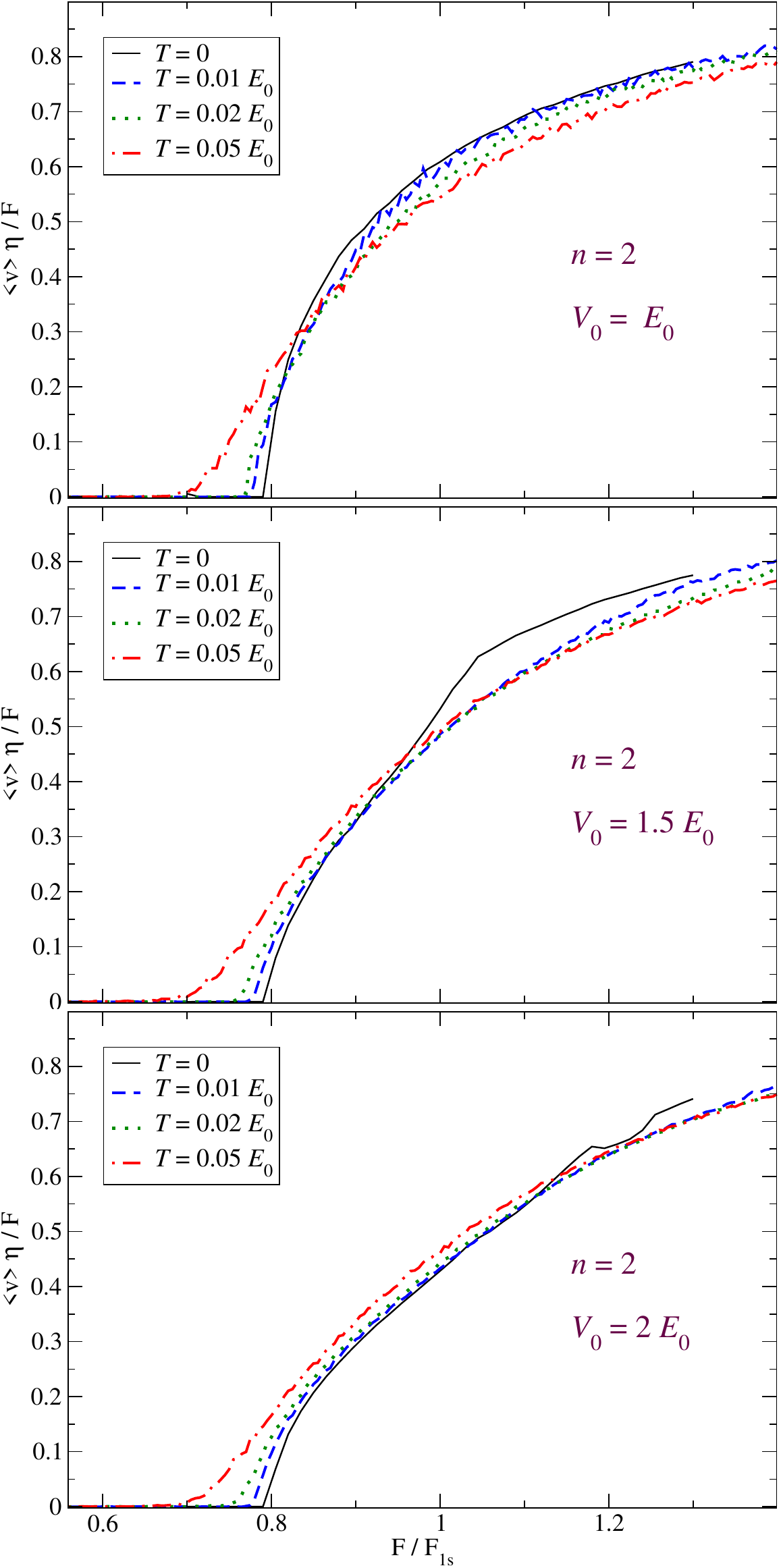}
}
\caption{\label{cl2:fig}
  The mobility $\langle v\rangle/F$, expressed in units of the free
  mobility $\eta^{-1}$, for clusters composed by $n=2$ particles.
  Simulations include $T=0$ and three finite but relatively small
  temperatures, expressed in units of $E_0$.
  The upper, central, and lower panels report the indicated values of the
  corrugation amplitude $V_0$, namely: $E_0$, $1.5\,E_0$, and $2\,E_0$
  respectively.
}
\end{figure}

\begin{figure}
\centerline{
\includegraphics[width=0.7 \columnwidth,clip=]{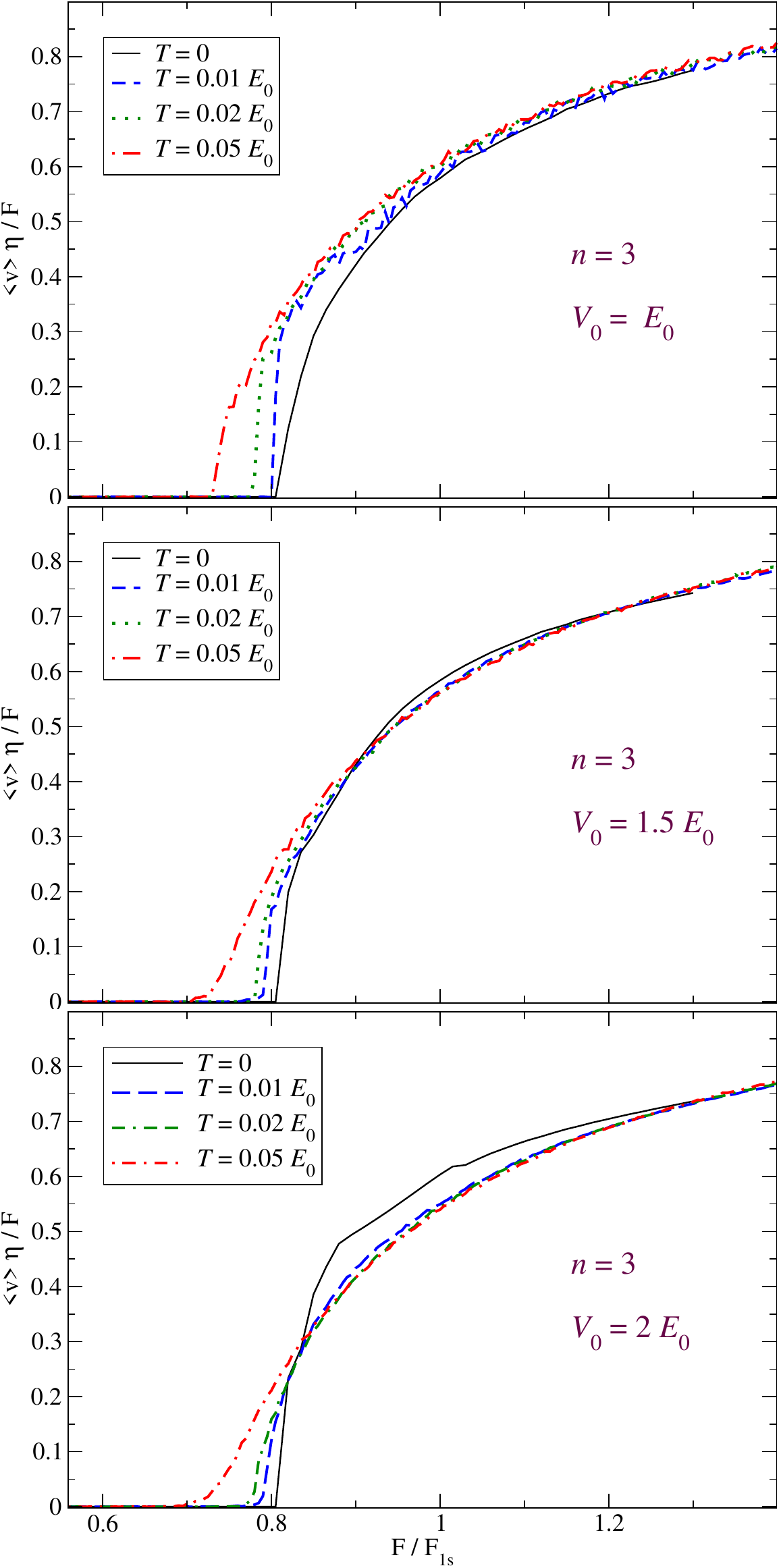}
}
\caption{\label{cl3:fig}
  Same as Fig.~\ref{cl2:fig}, but for clusters formed by $n=3$ particles.
}
\end{figure}

\begin{figure}
\centerline{
\includegraphics[width=0.7 \columnwidth,clip=]{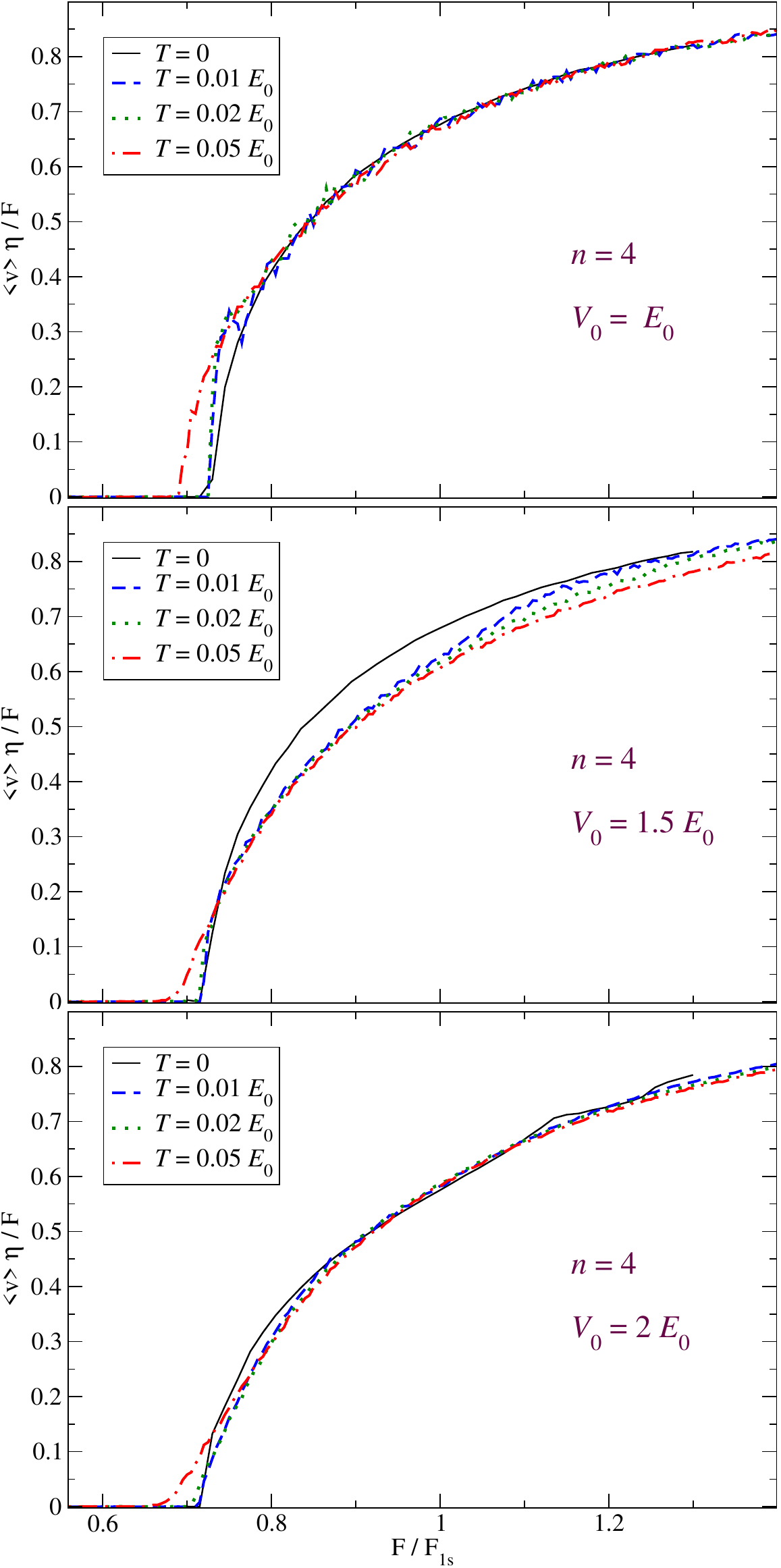}
}
\caption{\label{cl4:fig}
  Same as Fig.~\ref{cl2:fig}, but for clusters formed by $n=4$ particles.
}
\end{figure}

\begin{figure}
\centerline{
\includegraphics[width=0.7 \columnwidth,clip=]{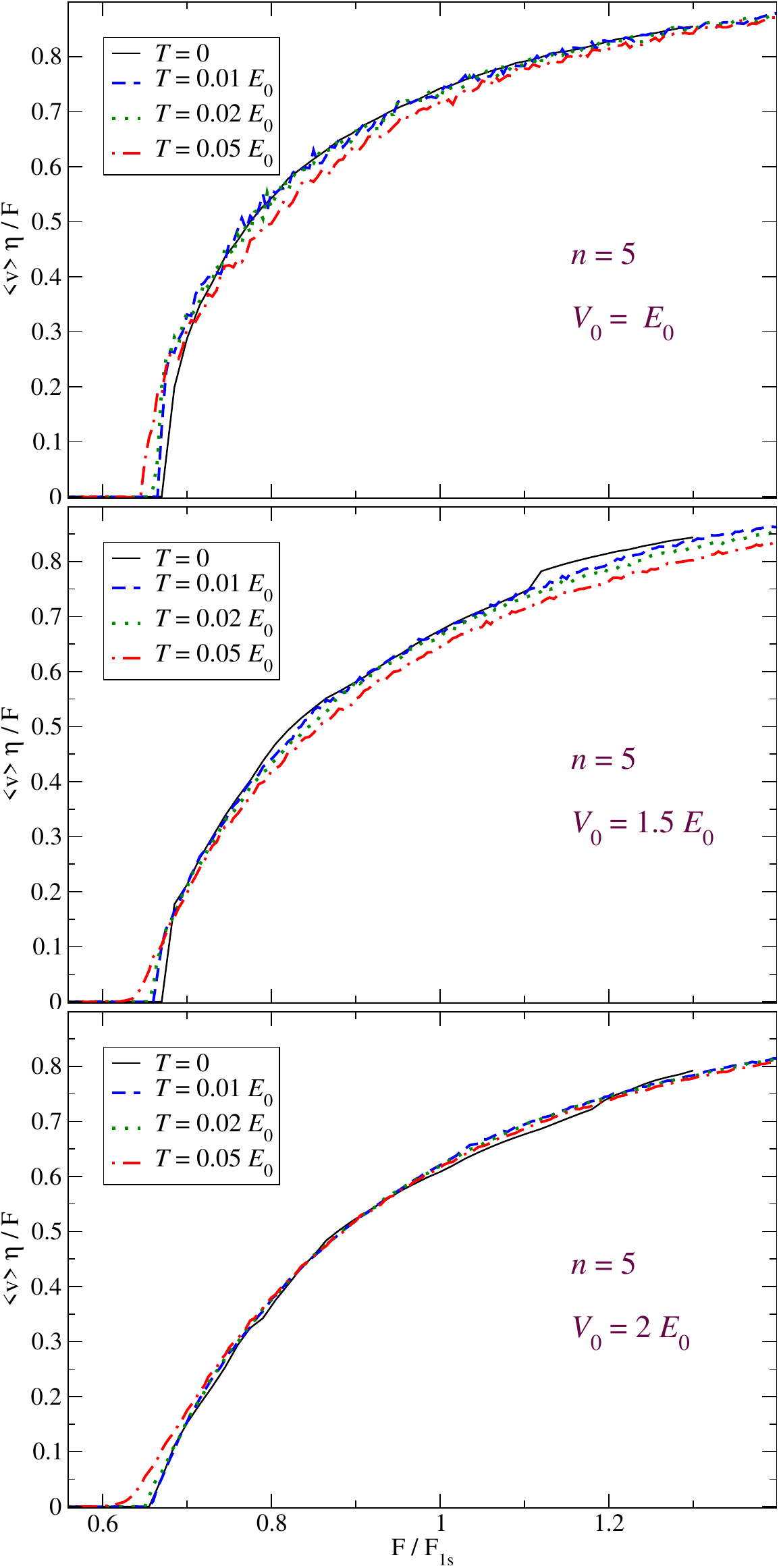}
}
\caption{\label{cl5:fig}
  Same as Fig.~\ref{cl2:fig}, but for clusters formed by $n=5$ particles.
}
\end{figure}

Figures \ref{cl2:fig} - \ref{cl5:fig} report simulated mobilities as a
function of the driving force.
The simulations are carried out at the values of the average particle
density given by Eq.~\eqref{rhoval:eq}, for $ n=2,3,4,5$, compatible with
clusters of $n$ particles arranged in a regular hexagonal lattice with
spacing $1.5\, R_c$.
Three different values of corrugation amplitude are considered, namely:
$V_0= E_0$, $V_0= 1.5\, E_0$, and $V_0= 2\, E_0$.
Each figure compares the $T=0$ mobility (also shown in the
Figs.~\ref{mob_cl2:fig}-\ref{mob_cl5:fig}) with the homologous quantity
obtained in finite-temperature simulations, carried out at three comparably
small temperatures: $T=0.01 \, E_0$, $T=0.02 \, E_0$, and $T=0.05\, E_0$.

%

\end{document}